\documentclass[lettersize,journal]{IEEEtran}
\usepackage{amsmath,amsfonts}
\usepackage{placeins}
\usepackage{algorithm}
\usepackage{algpseudocode}
\usepackage{array}
\usepackage[caption=false,font=normalsize,labelfont=sf,textfont=sf]{subfig}
\usepackage{textcomp}
\usepackage{stfloats}
\usepackage{url}
\usepackage{verbatim}
\usepackage{graphicx}
\usepackage{amssymb}
\usepackage{cite}
\usepackage{mathtools}
\usepackage{multicol}
\usepackage{multirow}
\usepackage{epstopdf}

\usepackage{ifsym}
\usepackage[shortlabels]{enumitem}
\usepackage{comment}
\usepackage{subfig}
\usepackage[table,xcdraw]{xcolor}

\makeatletter
\renewcommand\subsubsection{\@startsection{subsubsection}{3}{\z@}%
  {-1.25ex\@plus -1ex \@minus -.2ex}%
  {0.5ex \@plus .2ex}%
  {\normalfont\normalsize}}
\makeatother

\makeatletter
\renewcommand{\Function}[2]{%
  \csname ALG@cmd@\ALG@L @Function\endcsname{#1}{#2}%
  \def\jayden@currentfunction{#1}%
}
\newcommand{\funclabel}[1]{%
  \@bsphack
  \protected@write\@auxout{}{%
    \string\newlabel{#1}{{\jayden@currentfunction}{\thepage}}%
  }%
  \@esphack
}
\makeatother

\begin{document}

\title{A Novel Cipher for Enhancing MAVLink Security: Design, Security Analysis,  and Performance Evaluation Using a Drone Testbed}

\author{Bhavya Dixit, Ananthapadmanabhan A., Adheeba Thahsin, Saketh Pathak, Gaurav S. Kasbekar, and Arnab Maity 

\thanks{B. Dixit, Ananthapadmanabhan A., and
 G. Kasbekar are with the Department of Electrical Engineering, Indian Institute of Technology (IIT) Bombay, Mumbai 400076, Maharashtra, India. A. Thahsin, S. Pathak, and A. Maity are with the Department of Aerospace Engineering, IIT Bombay. Their email addresses are bdixit9@gmail.com, anantha9102002@gmail.com, gskasbekar@ee.iitb.ac.in, adeebathahsin@gmail.com, saketpathak321@gmail.com,  and arnab.maity@iitb.ac.in, respectively. The contributions of B. Dixit, A. Thahsin,  S. Pathak, G. Kasbekar, and A. Maity have been  supported in part by the project with code RGSTC01-001. A preliminary version of this paper was presented at the 39th International Conference on Information Networking (ICOIN) 2025  \cite{adheebaT}. }}

\maketitle

\begin{abstract}
Unmanned Aerial Vehicles (UAVs)  are being widely deployed for diverse applications, including surveillance, agriculture, logistics, disaster response, etc. Secure communication between a UAV and its ground control station (GCS) is paramount, as vulnerabilities can expose the system to cyber threats. Micro Air Vehicle Link (MAVLink) is a widely used open-source communication protocol that facilitates the exchange of messages between a UAV and a GCS. However, under this protocol, a UAV and a GCS communicate via an unencrypted channel, rendering the exchange of information vulnerable to eavesdropping attacks, and potentially compromising the mission objectives and sensitive information. Numerous research efforts have addressed the encryption of MAVLink protocol messages. However, most of these studies are either theoretical or based on simulations, rather than practical implementations. In this paper, we integrate various existing encryption algorithms, viz., Advanced Encryption Standard - Counter mode (AES-CTR), ChaCha20, Speck-CTR, and Rabbit, into MAVLink. We propose a novel cipher, MAVShield, designed to safeguard MAVLink-based communications. We perform a security analysis of MAVShield, which includes the study of $24$ distinct attacks on the proposed cipher using various statistical test suites, viz., the National Institute of Standards and Technology (NIST) and Diehard test suites. Our analysis demonstrates the robust resistance of MAVShield to differential cryptanalysis. Also, we thoroughly evaluate the performance of all five algorithms, viz., AES-CTR, ChaCha20, Speck-CTR, Rabbit, and MAVShield, and compare it with that of the standard unencrypted MAVLink protocol in terms of various metrics such as memory usage, battery power consumption, and CPU load, using a real drone testbed. Our performance evaluation demonstrates that MAVShield outperforms all the other encryption algorithms, and is hence a secure and efficient solution for protecting MAVLink-based communications in real-world deployments.
\end{abstract}


\section{Introduction} \label{section_1_intro}
\IEEEPARstart{A}{s} the skies become increasingly crowded with drones, the imperative to establish robust and reliable communication between Unmanned Aerial Vehicles (UAVs) and their ground control stations (GCS) has never been more critical \cite{chao2010autopilots}. UAVs have been extensively deployed in several military and civilian applications, ranging from disaster management to agriculture automation and from border surveillance to traffic monitoring. UAVs are entities that can be controlled remotely or fly autonomously using pre-programmed flight plans. With their ability to navigate complex landscapes and perform tasks without human intervention, UAVs are revolutionizing aerial operations \cite{chao2010autopilots}. In detail, an Unmanned Aerial System is composed of the following key elements \cite{pizzolante2023improving}:
\begin{itemize}
    \item 
    UAV: It comprises hardware and software that allow for flight control, navigation, and stabilization.
    \item 
GCS: It works as a central hub for monitoring drone status, controlling its movements, and receiving mission-specific data such as sensor readings. 
\item 
Communication links: Communication between the GCS and the UAV primarily relies on two links: the command $\&$ control link and the data link. The GCS sends instructions to the drone via the former link and receives vital telemetry data, such as position, speed, and battery status via the latter link.
\end{itemize}

Micro Air Vehicle Link (MAVLink) is a lightweight message serialization protocol, which has been specified to ensure efficient, reliable, and extensible communications between a UAV and a GCS \cite{koubaa2019micro}. It carries information about the status of the UAV as well as commands for control from the GCS. Lorenz Meier released MAVLink in 2009 under the GNU Lesser General Public License (LGPL) \cite{mavlinkio}. MAVLink's binary serialization feature produces smaller messages and incurs less overhead compared to other serialization techniques, contributing to its lightweight nature \cite{koubaa2019micro}. 

Security of UAV-GCS communication is paramount, as vulnerabilities can expose the system to cyber threats \cite{pizzolante2023improving}. 
However, although MAVLink is a robust communication protocol, it lacks adequate security mechanisms \cite{khan2022secure}.  In its earlier version, MAVLink 1.0, the protocol lacked native support for authentication and
authorization and was reported in the National Vulnerability Database \cite{nvdwebsite}. The current version, MAVLink 2.0, maintains backward compatibility with version 1.0 while introducing a key security feature: packet signing for authentication. This allows MAVLink systems to verify that messages originate from trusted sources. The signature is a 48-bit value derived from the first 48 bits of a SHA-256 hash, which is created from the entire packet combined with a secret key \cite{msgsign}. Despite these improvements, the protocol remains susceptible to eavesdropping attacks \cite{Xu}. As UAV-GCS communications occur over an unencrypted channel under MAVLink, they can be intercepted and read by malicious actors, leading to breaches in the confidentiality of the communication link \cite{hamza2024mavlink}.

To address these confidentiality threats, extensive research has been conducted on improving the security of MAVLink by integrating encryption into the communication link between a UAV and a GCS 
(see Section II for a review). However, most of these efforts primarily focused on theoretical frameworks and simulation-based evaluations. There has been a significant lack of real-world demonstrations showcasing the implementation of encryption within the MAVLink protocol as used in operational drones. In order to address this gap, in this paper, we integrate four encryption algorithms, viz., 
Advanced Encryption Standard - Counter mode (AES-CTR) \cite{yustiarini2022comparative},  ChaCha20 \cite{sabuwala2024approach}, Speck in Counter mode (Speck-CTR) \cite{beaulieu2017notes}, and Rabbit \cite{boesgaard2008rabbit} with the MAVLink protocol to ensure message confidentiality, and evaluate their performance using a hardware testbed. In addition, we propose a novel encryption algorithm, \emph{MAVShield}, integrate it with MAVLink, and show, via extensive experimentation using our testbed, that it significantly outperforms existing ciphers. To the best of our knowledge, this paper is the first  to integrate  encryption algorithms with MAVLink and evaluate their performance using a real drone testbed. 

Specifically, the contributions of this paper are as follows:
\begin{itemize}
    \item First, we integrate the existing encryption algorithms AES-CTR, ChaCha20, Speck-CTR, and Rabbit into MAVLink. 
    \item Second, we propose a novel cipher, MAVShield, designed to safeguard MAVLink based communications.
    \item Third, we thoroughly evaluate the performance of all five algorithms, viz. AES-CTR, ChaCha20, Speck-CTR, Rabbit, and MAVShield, and compare it with that of the standard unencrypted MAVLink protocol in terms of various metrics such as memory usage, battery power consumption, and CPU load, using a real drone testbed. The drone we use  is lab-engineered, and our testbed also includes a QGroundControl ground station \cite{knuthwebsite}, an ArduPilot  autopilot system \cite{ardu}, a Pixhawk Cube Orange$^{+}$ flight controller \cite{cube}, a Radiomaster boxer radio controller \cite{RC}, a telemetry radio module, a Global Positioning System (GPS) module, and a battery. Our evaluation demonstrates that MAVShield surpasses all the other encryption algorithms, and results in only marginal increases in memory usage ($0.024\, \%$), battery power consumption ($5.653 \, \%$), and CPU load ($2.937 \, \%$) relative to the unencrypted MAVLink protocol. These results confirm the efficiency of MAVShield and show that it is a feasible solution for use in practical UAV communication.
    \item Fourth, we perform a security analysis of MAVShield to assess its resilience against various attack vectors and its effectiveness in enhancing the overall security of UAV communications. Our analysis includes a study of $24$ distinct attacks on the proposed cipher using various statistical test suites, viz., the National Institute of Standards and Technology (NIST) \cite{rukhin2001statistical} and Diehard \cite{bogos2022remark} test suites, which demonstrates its robust resistance to differential cryptanalysis \cite{biham1991differential}. It shows that MAVShield can ensure the confidentiality of the secret key and plaintext data by preventing intruders from detecting statistical differences between the encryptions of two different messages, thereby validating the algorithm's security. 
\end{itemize}
Our results significantly enhance our understanding of MAVLink security as well as promote the development of strong countermeasures to safeguard the confidentiality of MAVLink-based UAV communications. Also, our experimental results using a real drone testbed and detailed security analysis establish that MAVShield is a feasible and effective solution for protecting MAVLink-based communications in real-world deployments.  

The rest of this paper is organized as follows. Section II reviews related work.  Section III presents the system model and problem formulation, core architecture of the MAVLink 2.0 protocol, its security weaknesses, and the adversary model. Section IV describes our encryption and decryption model and various existing encryption algorithms. Section V describes the proposed MAVShield cipher. Section VI provides a security analysis of MAVShield using various statistical suites. Section VII describes the experimental setup for our drone testbed. Section VIII explains how the encryption algorithms are implemented  in our testbed. Detailed experimental results, including a comparison of the performance of various existing encryption algorithms and the MAVShield algorithm, are provided in Section IX. Finally, Section X provides conclusions and directions for future research. 

\section{Related Work} \label{section2_lit_survey}

Ensuring security is crucial for the wide adoption of UAV systems. In this context, we now examine prior research on attacks on MAVLink communication links and the integration of encryption algorithms with the MAVLink protocol.

In \cite{allouch2019mavsec}, the authors have proposed a solution called MAVSec to address confidentiality and privacy problems in MAVLink. Various security algorithms, viz., AES-CTR, AES-Cipher Block Chaining (CBC), ChaCha20, and Rivest Cipher 4 (RC4), were implemented  on both an autopilot and a GCS. The authors compared the performances of different security solutions using an ArduPilot Simulation in the Loop (SITL) setup. A new solution has been proposed in \cite{pizzolante2023improving} to address the problem of key exchange between the UAV and the GCS. Also, the authors have compared the performances of fourteen encryption algorithms using a virtual drone and determined the best algorithm to be Speck 128/192, which was later used with a key exchange phase. They concluded that this causes only a slight overhead relative to using only encryption. In \cite{khan2022secure}, a new approach based on an  encryption algorithm and a custom mapping has been applied to secure the MAVLink communication protocol. In this approach, a new concept of lists has been introduced, under which a serial number is shared instead of directly sending the secret key used for encryption between the UAV and GCS, thereby ensuring that the actual secret key is not exposed during transmission. The proposed method is compared with the original MAVLink protocol in a simulation environment and the authors get almost similar results in terms of transfer speed, memory consumption, and CPU usage for the two protocols. In \cite{kassim2022dmav}, the authors have proposed an innovative method called DMAV based on dynamic DNA coding  to strengthen the security of the MAVLink protocol. Their approach encrypts the MAVLink payload using DNA coding with binary bits and the lightweight GIFT algorithm. The technique has been compared with the original, unsecured MAVLink protocol in the ArduPilot-SITL environment, demonstrating that it effectively secures communication without introducing much system overhead. In another study \cite{sabuwala2022securing}, the authors have secured the MAVLink-based communication between the GCS and the UAV through implementation of the following security algorithms: encryption by Navid \cite{khan2022secure}, DMAV \cite{kassim2022dmav}, and ChaCha20 \cite{bernstein2008chacha}. In a performance evaluation conducted in the Gazebo simulation environment, ChaCha20 emerged as the most efficient algorithm, outperforming the others in terms of memory consumption, CPU utilization, and packet transfer rate. Further research contributions in \cite{sabuwala2024approach} proposed integrating a novel stream cipher into the MAVLink communication protocol. The authors have designed the new cipher by modifying the Quarter Round function of the ChaCha20 algorithm to enhance the security without causing much overhead in the system. In \cite{manesh2019cyber}, the authors have provided an overview of data interception attacks, data manipulation attacks, and denial-of-service attacks on four different channels, viz., UAV-GCS, UAV-GPS, UAV-other aircraft, and UAV-automatic dependent surveillance-broadcast (ADSB) system. Also, defense strategies for several attacks have been discussed and compared.

In the above studies \cite{pizzolante2023improving,khan2022secure,allouch2019mavsec,kassim2022dmav,sabuwala2022securing,sabuwala2024approach,manesh2019cyber}, researchers have proposed security protocols for encrypting MAVLink and evaluated their effectiveness via simulations; however, none of these studies have implemented these security measures in a real drone testbed or tested their performance in real-world conditions.
The practicality and safety of implementing their solutions in actual scenarios remain uncertain, as their assessments were confined to simulation environments. In contrast, in this paper, we implement several existing encryption algorithms using a real drone testbed, propose a new encryption method, as well as compare its performance with that of the existing security schemes using our testbed.

In \cite{fanfakh2024parallel}, the authors have proposed an optimized version of the Speck cipher that leverages a substitution table to enhance the efficiency of the encryption process. By updating the S-box in each iteration using the RC4 algorithm, they were able to reduce the number of rounds required for the Speck 128/128 cipher. Building on this concept, this paper presents a novel encryption approach, MAVShield, which enhances the Speck round function by dynamically updating the key value. This is accomplished through a sophisticated process, which involves word splitting, S-box substitution, and XOR operations. Furthermore, by incorporating a nonce, the same approach is utilized for key generation after applying rotation and arithmetic operations to the nonce, ensuring that the secret key for each round is completely random. Our experimental findings from the drone testbed reveal that MAVShield surpasses Speck-CTR, as well as AES-CTR, ChaCha20, and Rabbit, in terms of battery power consumption, memory usage, and CPU utilization.

\section{System Model And Problem Formulation} \label{section3}
This section describes the system model in which UAV-GCS communication takes place using MAVLink, which lacks built-in confidentiality measures. Also, it describes the objective of this paper, which is to  evaluate existing encryption algorithms using a drone testbed and to design a new cipher, viz., MAVShield. It also describes the adversary model, which highlights cyber threats such as eavesdropping and data interception, and stresses the need for robust encryption.

\begin{figure}[h]
    \centering
    \includegraphics[width=0.7\linewidth]{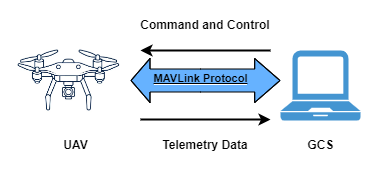}
    \caption{The figure shows the communication between a UAV and a GCS.}
    \label{D1}
\end{figure}

\subsection{System Model}
In this study, we focus on a communication link employing the MAVLink protocol between a UAV and a GCS. Here, communication takes place over a wireless link, which commonly utilizes a radio frequency (RF) link or the Wi-Fi protocol. 

The MAVLink protocol offers bidirectional communication between a GCS and a UAV (see Fig. \ref{D1}). The channel from the GCS to the UAV is used to send command and control messages, while the channel from the UAV to the GCS is used for telemetry data transmission such as payload data and status information. To keep the communication alive, the UAV periodically transmits a HEARTBEAT message \cite{koubaa2019micro} to the GCS.

\begin{figure*}
    \centering
    \includegraphics[width=\textwidth]{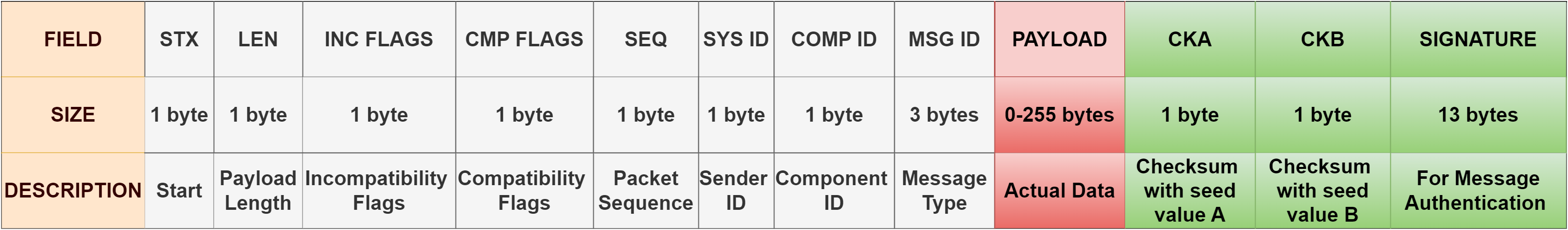}
    \caption{The figure shows the MAVLink 2.0 packet structure. \iffalse{\bf (DONE: Change ``FIELDS'' to ``FIELD''. Also, should ``0xFD'' be replaced with the size in bytes?)}\fi}
    \label{D2}
\end{figure*}

\begin{figure*}[ht]
    \centering
    \includegraphics[width=\textwidth]{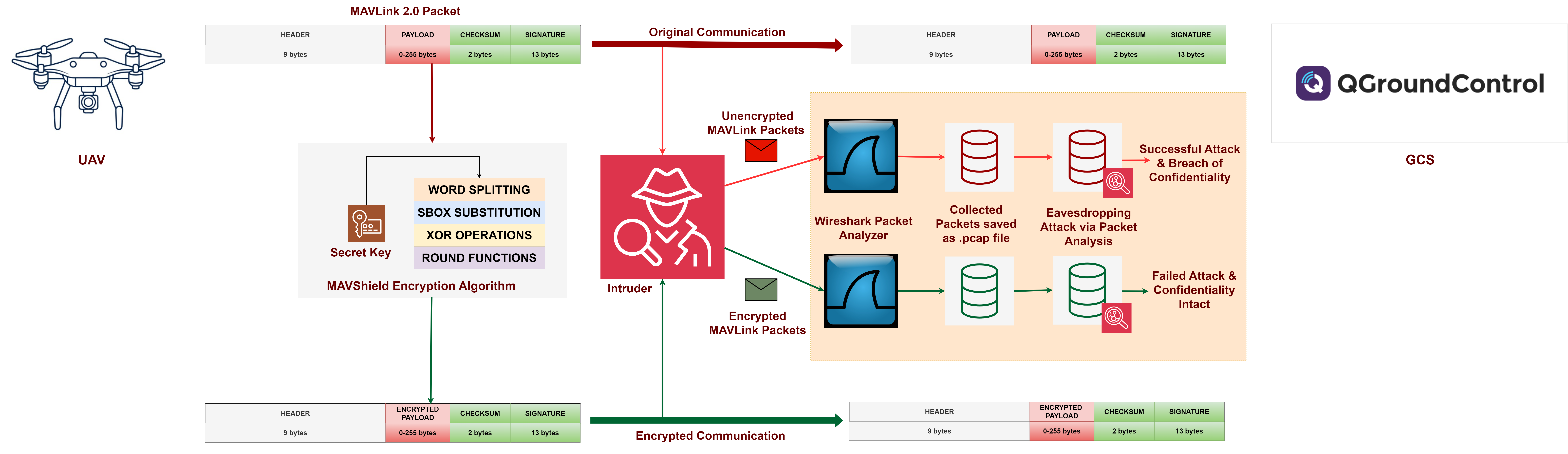}
    \caption{The figure shows the adversary model.}
    \label{D4}
\end{figure*}

MAVLink is a lightweight and open-source message serialization protocol \cite{sabuwala2022securing}. It is mainly used for two-way information exchange between a UAV and a GCS. It is designed to be lightweight because it is used for real-time communication. 

Additionally, MAVLink messages are serialized at the application layer before being passed to the lower layers. Their transmission can occur over various networks and protocol stacks, including the TCP/ IP protocol stack, Wi-Fi (which operates at the  medium access control (MAC) and physical layers), and low-bandwidth serial telemetry links \cite{sabuwala2022securing}. This serialization and de-serialization process occurs at both ends of the communication link and minimizes the number of transmission messages required for the serialization. As a result, the design is well-suited for resource-constrained environments.

Fig. \ref{D2} shows the MAVLink 2.0 packet structure \cite{koubaa2019micro}. Message types are identified by message IDs and the ``payload'' field contains the actual data. A MAVLink message has a packet size of 11 to 297 bytes, of which the maximum length of the actual payload data is 255 bytes. Each MAVLink message includes  a few error detection and security mechanisms \cite{kurose2005computer}:
\begin{itemize}
    \item {\bf Checksum:} It facilitates the detection of message corruption, serving as an error detection mechanism to ensure the correctness of data transmitted over a network.
    \item {\bf Signature:} It enables message authentication, i.e., verification that the message comes from a trusted source and has not been modified during transit.
\end{itemize}
The inclusion of a signature in a MAVLink packet plays a crucial role in preventing the tampering or spoofing of messages during transmission. However, despite these security measures, an adversary with access to the channel can still eavesdrop on the transmitted messages, observing the communication flow without altering its contents.

\subsection{Objectives}

MAVLink messages exchanged over the communication link are not encrypted, and an attacker can intercept the data and extract sensitive information. Therefore, to address this vulnerability of the MAVLink protocol and to achieve true security, encryption is necessary. Our objective is to strengthen the protocol's security by integrating encryption into it, while preserving its existing integrity and authentication features. We seek to integrate existing encryption algorithms such as AES-CTR, Speck-CTR,  ChaCha20, and Rabbit into MAVLink and evaluate their performance in terms of various performance metrics such as memory usage, CPU utilization, and battery power consumption by conducting experiments on a real drone testbed. Another goal is to design a new encryption algorithm that outperforms existing ones, evaluate its performance on our drone testbed, and conduct its thorough security analysis.

\subsection{Security Issues and Adversary Model}

In general, attacks on networks can compromise the confidentiality, integrity, availability, and authenticity of the system \cite{perlman2016network}. They can result in interception (compromising data confidentiality and privacy), modification (compromising data integrity), interruption (compromising data availability), and/ or fabrication (attacks on authenticity).

Confidentiality means that only authorized users should be able to read transmitted messages; if it is violated, then an attacker can eavesdrop on the channel and read messages such as sensor readings, GPS  data, telemetry feeds, and commands communicated from the GCS to the UAV \cite{kwon2018empirical}. Recall that the MAVLink protocol incorporates security mechanisms for achieving message integrity and authentication, but not for achieving confidentiality.  To mitigate this weakness and to protect data in transit, we implement encryption mechanisms, thereby securing the communication channel.

The adversary model shown in Fig. \ref{D4} illustrates both the vulnerabilities that exist in data transmission and the defensive measures necessary to maintain confidentiality within the system.  In particular, the upper part of the figure shows that the UAV transmits MAVLink packets to the GCS software, QGroundControl, via a communication channel. An intruder monitoring the transmission can capture the data, e.g., in a packet capture (.pcap) file using a Wireshark packet analyzer. Since the messages are in clear text, they can be easily intercepted and analyzed, leading to a breach of confidentiality.  The lower part of the figure shows the proposed solution, in which MAVLink packets are encrypted using the MAVShield encryption algorithm. This method involves word splitting, S-box substitution, and XOR operations of 64-bit secret key, followed by the application of the updated key to the round functions. If an adversary intercepts the message transmitted over this encrypted channel and attempts to collect and analyze the data, they fail since the message is unintelligible to them. Consequently, confidentiality remains intact. Thus, this model highlights the importance of robust encryption techniques in protecting UAV communications against potential threats from intruders. 

\section{Security in MAVLink Communications}
\label{section4}

Since under MAVLink 2.0, UAV-GCS communications are vulnerable to attacks against data confidentiality and privacy,  we encrypt the communication channel. In this section, we describe our encryption and decryption model and different existing encrytion algorithms, viz., AES, Speck,  ChaCha20, and Rabbit, which we integrate into MAVLink. 

\subsection{Encryption and Decryption Model }\label{section4.1} 

\begin{figure}[!t]
\centering
\subfloat[\scriptsize{MAVLink payload encryption}]{\includegraphics[width=0.47\linewidth]{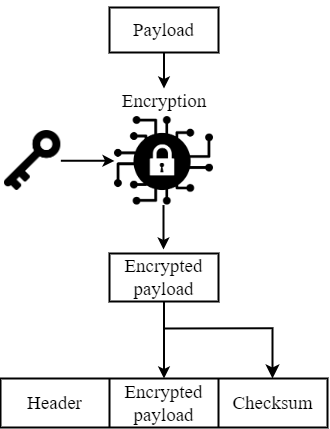}%
\label{fig_encryption}}
\hfil
\subfloat[\scriptsize{MAVLink payload decryption}]{\includegraphics[width=0.52\linewidth]{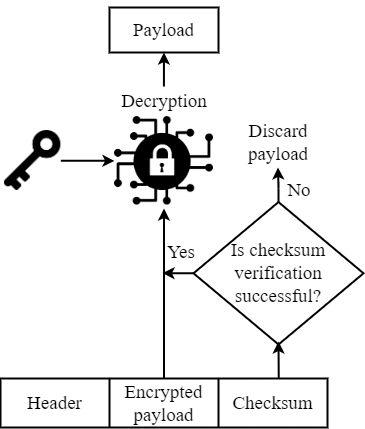}%
\label{fig_decryption}}
\caption{The figure shows the integration of security algorithms in the MAVLink protocol.}
\label{D5}
\end{figure}

Fig. \ref{D5} illustrates the integration of an encryption algorithm within the MAVLink protocol. To preserve the proper functionality of MAVLink communication, at the transmitter, encryption is applied exclusively to the payload data, as depicted in Fig. \ref{fig_encryption}. Encryption of the header is not done since it could prevent the recipient from identifying the message type \cite{koubaa2019micro}. Consequently, MAVLink packets retain an unencrypted MAVLink ID in the header to ensure seamless message recognition. Furthermore, recall that MAVLink incorporates a checksum mechanism for the detection of bit errors; under our model,  encryption is performed before checksum computation, thereby safeguarding the confidentiality of the payload.

Subsequently, at the receiver, as depicted in Fig. \ref{fig_decryption}, decryption occurs only after successful checksum verification; if the checksum fails, the payload is discarded. Hence, our encryption-decryption process guarantees the confidentiality and functionality of MAVLink communication.

\subsection{Encryption Algorithms  }\label{section4.2}  

We integrate the security algorithm  into both ends of the MAVLink  communication channel, i.e., in the UAV and GCS. To address the confidentiality flaws in MAVLink, we use four different  symmetric key cryptography algorithms, viz., AES-CTR, ChaCha20, Speck-CTR, and Rabbit. Here, ChaCha20 and Rabbit are stream ciphers, while AES and Speck are block ciphers \cite{perlman2016network}. 

Block ciphers support various modes of operation, including Electronic Codebook (ECB), CBC, Output Feedback (OFB), and CTR, each tailored to specific use cases \cite{perlman2016network}. Among these, CTR mode is particularly well-suited for encrypting MAVLink payloads, because in this mode, a one-time pad (OTP) can be pre-computed, allowing encryption to be performed by simply XORing the OTP with the plaintext once the latter becomes available. This approach reduces the encryption latency compared to other modes, such as ECB and CBC \cite{perlman2016network}. Another advantage of the CTR mode is that it accommodates arbitrary payload lengths and maintains a ciphertext length equal to the plaintext length. While the CBC mode is also compatible with the considered block ciphers, AES and Speck, its use has been constrained by challenges in supporting arbitrary payload lengths \cite{pizzolante2023improving}. 

\begin{figure}[h]
    \centering
     \includegraphics[width=0.98\linewidth]{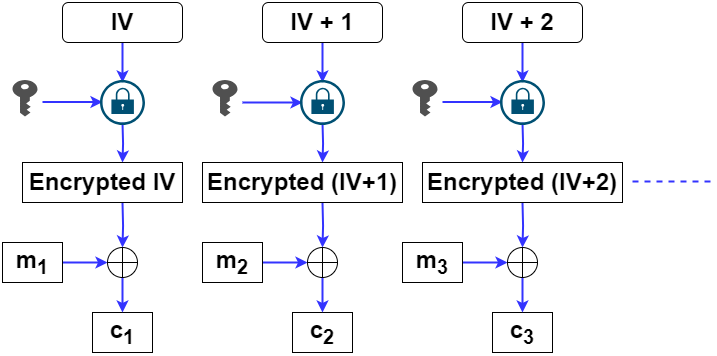}
    \caption{The figure shows the encryption process in counter mode \cite{perlman2016network}. \iffalse{\bf (DONE: Add $\ldots$ on the right of the figure.)}\fi}
    \label{ctr}
\end{figure}

As shown in Fig. \ref{ctr}, CTR mode converts a block cipher into a stream cipher. The considered block ciphers  employ a randomly generated 128-bit initialization vector (IV), also known as the initial counter, which is unique for each encryption process. For each plaintext block, the counter value is incremented and then encrypted using the secret key to generate a keystream block. The ciphertext block is then obtained by XORing the plaintext block with the corresponding keystream block. Decryption follows the same process, where the ciphertext block is XORed with the keystream block generated using the same counter value to retrieve the original plaintext. Since encryption and decryption are independent for each block, they can be performed in parallel. Additionally, this independence makes the CTR mode resilient to block loss \cite{perlman2016network}.

\subsubsection{Advanced Encryption Standard (AES)}\label{section4.2.1}  

\begin{figure}[h]
    \centering
     \includegraphics[width=0.8\linewidth]{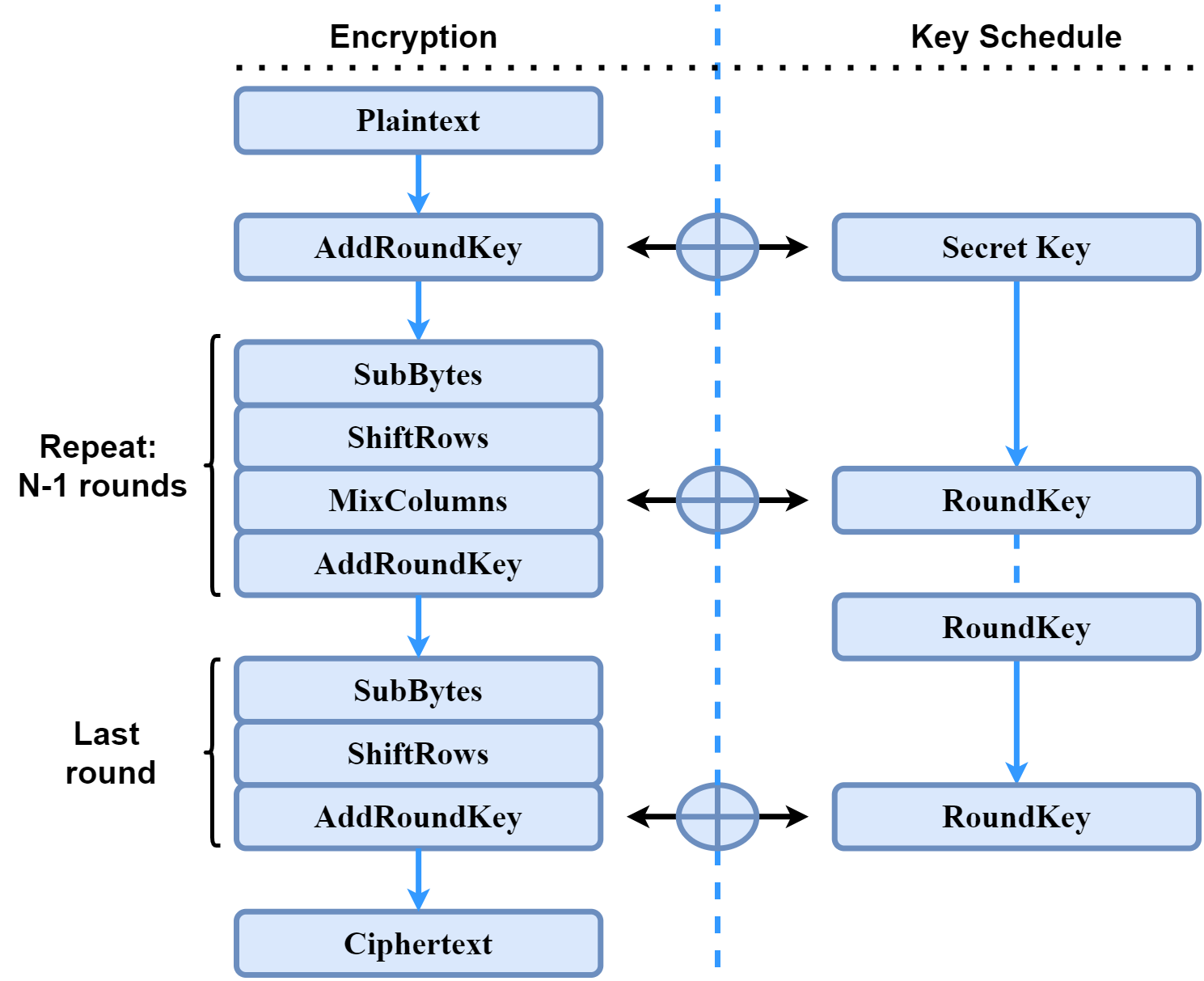}
    \caption{The figure shows the encryption process of the AES algorithm. \iffalse{\bf (DONE: Change ``Repeat $N-1$ round'' to ``Repeat: $N-1$ rounds''.) }\fi}
    \label{D6}
\end{figure}

AES employs a symmetric key algorithm, utilizing the same secret key for encryption and decryption. The AES algorithm uses substitutions, sequences, and a series of rounds applied to each encrypted and decrypted block. This algorithm has 128-bit, 192-bit, or 256-bit long keys \cite{yustiarini2022comparative}. It is a highly secure and efficient method of encryption. It was established by the U.S. NIST in 2001 and has since become a globally accepted encryption standard. 

As shown in Fig. \ref{D6}, AES employs a unique internal key, known as a ``round key'', derived from the secret key, for each round of encryption. This algorithm involves four types of byte transformations: SubBytes, ShiftRows, MixColumns, and AddRoundKey \cite{yustiarini2022comparative}. At the start of the first round, the plaintext is copied into the state and then  undergoes the byte transformation AddRoundKey. After that, the state is transformed by the SubBytes, ShiftRows, MixColumns, and AddRoundKey operations $N -1$ times, where $N$ is the total number of rounds   \cite{yustiarini2022comparative}. This set of operations is called the AES round function. In the last round, the state is transformed by the SubBytes, ShiftRows, and AddRoundKey operations and it does not experience the MixColumns transformation. This iterative approach in AES enhances security by using a complex encryption process.

\subsubsection{ChaCha20}\label{section4.2.2}

The ChaCha family of stream ciphers is a category of high-throughput stream ciphers that are primarily designed for software platforms. It achieves a good balance between security and performance \cite{sabuwala2024approach}.
Algorithm \ref{alg4} provides the pseudo-code for ChaCha20, detailing its steps and structure \cite{sabuwala2024approach}.

\begin{algorithm}
\caption{Pseudo-code of ChaCha20 Algorithm}\label{alg4}
\textbf{Input: $Key \in {\{0,1\}}^{256}$, $Nonce \in {\{0,1\}}^{96}$ , $Count \in {\{0,1\}}^{32}$ , $PlainText \in {\{0,1\}}^*$} \\
\textbf{Output: $CipherText = ChaCha20(Key, Nonce, Count, $\newline$ PlainText)$} 
\begin{algorithmic}[1]
\State $I \gets Init(Key, Nonce, Count)$ 
\For{$a \gets 1$ to $(\frac{no\_of\_bits\_PlainText}{512})$} \Comment{(a is block number)}
    \State $O \gets I$
    \For{$b \gets 1$ to $10$}  
    \State $O[0,4,8,12] \gets QR(O[0,4,8,12])$
    \State $O[1,5,9,13] \gets QR(O[1,5,9,13])$
    \State $O[2,6,10,14] \gets QR(O[2,6,10,14])$
    \State $O[3,7,11,15] \gets QR(O[3,7,11,15])$
    \State $O[0,5,10,15] \gets QR(O[0,5,10,15])$
    \State $O[1,6,11,12] \gets QR(O[1,6,11,12])$
    \State $O[2,7,8,13] \gets QR(O[2,7,8,13])$
    \State $O[3,4,9,14] \gets QR(O[3,4,9,14])$
    \EndFor
    \State $Sl \gets Serial(O + I)$ \Comment{(Serialized keystream)}
           \For{$c \gets 1$ to $512$}
           \State $CipherText[512(a-1) + (c-1)] \gets PlainText[512(a-1) + (c-1)] \oplus Sl[c-1]$
           \EndFor
    \State $I[12] \gets I[12] + 1$  \Comment{(Counter value incremented by one)}      
 \EndFor
  \State \Return $CipherText$
\end{algorithmic}
\end{algorithm}

The ChaCha20 algorithm begins with an initial row-vector $I$, which contains four $32$-bit constants, a $256$-bit key, a $32$-bit initial counter, and a $96$-bit nonce. This vector is structured as a $4 \times 4$ grid, with each entry representing a $32$-bit word. The initial vector $I$ is then utilized to create a $512$-bit operational vector $O$. In ChaCha20, the internal state $O$ undergoes $10$ double rounds, resulting in a total of $20$ updates. Each round  consists of four quarter rounds (QR). The quarter round applies four operations to four state variables, mixing the data through addition, bitwise XOR, and rotation operations. The rounds conclude once every state in $O$ has been updated. The final output of $O$ is combined with $I$. The addition operator, $+$, combines the current state value $O$ with the initial value $I$, generating the keystream and serializing $32$-bit values into a $512$-bit keystream. The plaintext is then divided into $512$-bit blocks, and the ciphertext is produced by performing an XOR operation between the plaintext blocks and the generated keystream \cite{sabuwala2024approach}. 

The process of decryption is similar to encryption. During decryption, we use the keystream generated in the same manner as during encryption, and XOR it with the ciphertext to recover the plaintext.

\subsubsection{Speck}\label{section4.2.3} 

Speck is a lightweight block cipher offering robust security and efficient performance in both hardware and software \cite{beaulieu2017notes}. Developed in 2013 by the U.S. National Security Agency (NSA), it is optimized for resource-constrained devices using streamlined cryptographic techniques.

\begin{figure}[h]
    \centering
     \includegraphics[width=0.98\linewidth]{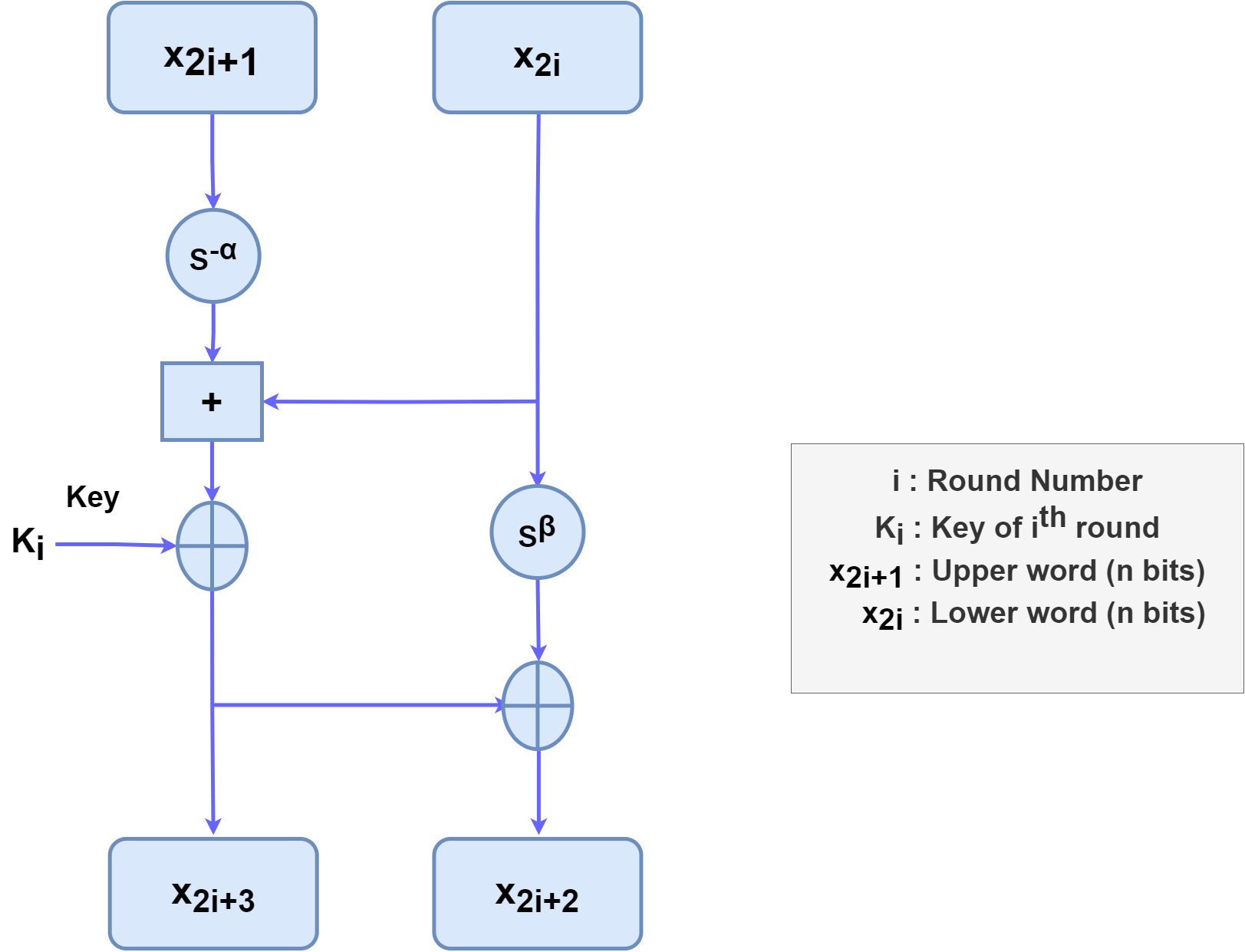}
    \caption{The figure shows the Speck round function \cite{sawant2019implementation}.}
    \label{D24}
\end{figure}

As illustrated in Fig. \ref{D24}, Speck is an Add-Rotate-XOR (ARX) block cipher with a Feistel-like structure, where both branches are updated in each round \cite{perlman2016network}. Its $2n$ block encryption process applies the following operations to each of two $n$-bit words: left or right circular shifts, modulo-2 addition, and bitwise XOR.  These operations collectively form the algorithm's round function. Non-linearity in its design comes from modular addition, making the algorithm strong cryptographically \cite{beaulieu2017notes}. In Fig. \ref{D24}, we use the following notation for operations on $n$-bit words \cite{beaulieu2015simon}:
\begin{itemize}
    \item bitwise XOR: $\oplus$,
    \item mod 2 addition: $+$,
    \item left circular shift by $\beta$ bits: $S^\beta$,
    \item right circular shift by $\alpha$ bits: $S^{-\alpha}$.
\end{itemize}
Also, $x_{2i+1}$ and $x_{2i}$ are the upper and lower words, respectively, of the state at the beginning of the $i^{th}$ round. The following two equations outline how $x_{2i+3}$ and $x_{2i+2}$ are computed:
\begin{align}
    x_{2i+3} &= ((S^{-\alpha}(x_{2i+1})) + x_{2i}) \oplus K_i, \\ 
    x_{2i+2} &= (S^{\beta}(x_{2i})) \oplus x_{2i+3}.     
\end{align}

Speck has multiple possible instantiations, supporting block sizes of 32, 48, 64, 96, and 128 bits, and with up to three key sizes to go along with each block size, as detailed in Table \ref{table:V}. The Speck family provides ten algorithms in all. In this paper, we  use a block size of 128  with key lengths of 128 bits, 192 bits, and 256 bits. For each key size, the number of rounds is different. For each round, a unique key is used for encryption. Round keys are expanded by key schedules. When Speck is operated in the CTR mode, the encryption and decryption processes described in Section \ref{section4.2} are used. 

\begin{table}[!ht]
\caption{The table shows the Speck parameters \cite{beaulieu2015simon}.}
\label{table:V}
\begin{tabular}{|l|l|l|l|l|}
\hline
\multicolumn{1}{|c|}{\cellcolor[HTML]{FFFFFF}{\color[HTML]{330001} \textbf{\begin{tabular}[c]{@{}c@{}}Block Size \\ 2n\end{tabular}}}} & \multicolumn{1}{c|}{\cellcolor[HTML]{FFFFFF}{\color[HTML]{330001} \textbf{\begin{tabular}[c]{@{}c@{}}Key Size \\ mn\end{tabular}}}} & \multicolumn{1}{c|}{\textbf{\begin{tabular}[c]{@{}c@{}}Keyword \\ m\end{tabular}}} & \textbf{$\alpha, \beta$} & \textbf{Speck Rounds} \\ \hline
\cellcolor[HTML]{FFFFFF}{\color[HTML]{330001} \textbf{32}}                                                                             & \cellcolor[HTML]{FFFFFF}{\color[HTML]{330001} \textbf{64}}                                                                          & \textbf{4}                                                                         & \textbf{7, 2}            & \textbf{22}           \\ \hline
\cellcolor[HTML]{FFFFFF}{\color[HTML]{330001} \textbf{48}}                                                                             & \cellcolor[HTML]{FFFFFF}{\color[HTML]{330001} \textbf{72, 96}}                                                                      & \textbf{3, 4}                                                                      & \textbf{8, 3}            & \textbf{22, 23}       \\ \hline
\textbf{64}                                                                                                                            & \textbf{96, 128}                                                                                                                    & \textbf{3, 4}                                                                      & \textbf{8, 3}            & \textbf{26, 27}       \\ \hline
\textbf{96}                                                                                                                            & \textbf{96, 144}                                                                                                                    & \textbf{2, 3}                                                                      & \textbf{8, 3}            & \textbf{28, 29}       \\ \hline
\textbf{128}                                                                                                                           & \textbf{128, 192, 256}                                                                                                              & \textbf{2, 3, 4}                                                                   & \textbf{8, 3}            & \textbf{32, 33, 34}   \\ \hline
\end{tabular}
\end{table}

\subsubsection{Rabbit}\label{section4.2.4} 

Rabbit is a stream cipher based on iterating a set of coupled non-linear functions \cite{boesgaard2003rabbit}. It takes a 128-bit secret key as input and generates an output block of 128 pseudo-random bits from a combination of internal state bits in each iteration. 

The internal state spans 513 bits, which are divided between eight 32-bit state variables, $x_{j, i}$, eight 32-bit counters,  $c_{j, i}$, and one counter carry bit, $\phi_{7, i}$, where $i,j \in \{0,\ldots,7\}$. Here, $x_{j, i}$ and  $c_{j, i}$ represent the variables of subsystem $j$ at iteration $i$. Each subsystem $j$ corresponds to one of the eight distinct state variables that form the cipher's internal state, while $i$ indicates the current iteration number. The eight state variables and counters are initialized using the key, while the counter carry bit starts at zero. The state variables are updated by eight coupled non-linear integer-valued functions \cite{boesgaard2003rabbit}. 

\paragraph{Key Setup Scheme}
This algorithm's initialization process involves expanding a 128-bit key into eight state variables and eight counters. The key is divided into segments and each segment is used to initialize one state variable $x_{j,0}$ and one counter $c_{j,0}$. 
The key, $K^{[127\ldots0]}$, is divided into eight subkeys: $k_0 = K^{[15\ldots0]}$, $k_1 = K^{[31\ldots16]}$, \ldots, $k_7 = K^{[127\ldots112]}$. 

Let the symbols $\diamond$ and $\oplus$ denote the concatenation of two bit sequences and the logical XOR operation, respectively.
The initial values of the state and counter variables are derived from subkeys as follows:
\begin{align}
    x_{j, 0} =
    \begin{cases}
        k_{((j+1) \text{ mod } 8)}  \diamond k_j, & \text{ for j even,}\\
         k_{((j+5) \text{ mod } 8)}  \diamond k_{((j+4) \text{ mod } 8)}, & \text{ for j odd,}
    \end{cases}
\end{align}
\begin{align}
    c_{j, 0} =
    \begin{cases}
        k_{((j+4) \text{ mod } 8)}  \diamond k_{((j+5) \text{ mod } 8)}, & \text{for j even,}\\
        k_j  \diamond k_{((j+1) \text{ mod } 8),} & \text{for j odd}.        
    \end{cases}
\end{align}

Based on the next-state function, the system undergoes four iterations, to reduce the correlation between the key bits and internal state variables. After these iterations, the counter values are updated as follows to protect the key from any potential recovery attempts through reversal of the counter system:
\begin{align}
    c_{j,4} = c_{j,4} \oplus x_{((j+4) \text{ mod }8), 4}.
\end{align}

\paragraph{Next State Function}
Let the notations $\lll$ and $\gg$ represent left bit-wise rotation and right logical bit-wise shift, respectively. The core of the Rabbit algorithm is the iteration of the system defined by the following equations \cite{boesgaard2008rabbit}:
\begin{align}
    g_{j, i} = ((x_{j, i} + c_{j, i})^2 \oplus ((x_{j, i} + c_{j, i})^2 \gg 32)) \text{ mod } 2^{32},    
\end{align}
\begin{align}
    x_{j, i+1} = 
     \begin{cases}
         g_{j,i}  + (g_{(j-1)\text{ mod }8, i} \lll 16) 
     \\
        +(g_{(j-2)\text{ mod }8, i} \lll 16), & \text{ if j is even,}\\
        g_{j,i} + (g_{(j-1)\text{ mod }8, i} \lll 8)  \\
        +(g_{(j-2)\text{ mod }8, i}), & \text{ if j is odd.}
     \end{cases}  
\end{align}
In these equations, all additions are performed modulo $2^{32}$. The term $g_{j, i}$ is the output of the non-linear function applied to the internal state variables during the $i^{th}$ iteration for subsystem $j$. 

\paragraph{Counter System}
\begin{align}
    c_{j, i+1} = 
     \begin{cases}
         (c_{0,i}  + a_0 + \phi_{7, i}) \text{ mod } 2^{32}, &       \text{      if } j = 0,\\
       (c_{j,i}  + a_j + \phi_{j-1, i+1}) \text{ mod } 2^{32}, & \text{ if } j  \in \{1,\ldots,7\},
     \end{cases}  
\end{align}
where the counter carry bit $\phi_{7, i}$ is given by:
\begin{align}
    \phi_{j, i+1} = 
     \begin{cases}
        1, & \text{ if } (c_{0,i}  + a_0 + \phi_{7, i} \geq 2^{32}) \land (j = 0), \\
       1, & \text{ if } (c_{j,i}  + a_j + \phi_{j-1, i+1} \geq 2^{32}) \land (j > 0), \\
       0, & \text{ otherwise. }
     \end{cases}  
\end{align}
Here, the constants, $a_0$, $a_3$, and $a_6$ are $0$x$4D34D34D$, $a_2$ and $a_5$ are $0$x$34D34D34$, and $a_1$, $a_4$, and $a_7$ are $0$x$D34D34D3$. The symbol $\land$ denotes the logical AND operation. 

\begin{figure*}[h]
    \centering
     \includegraphics[width=1\linewidth]{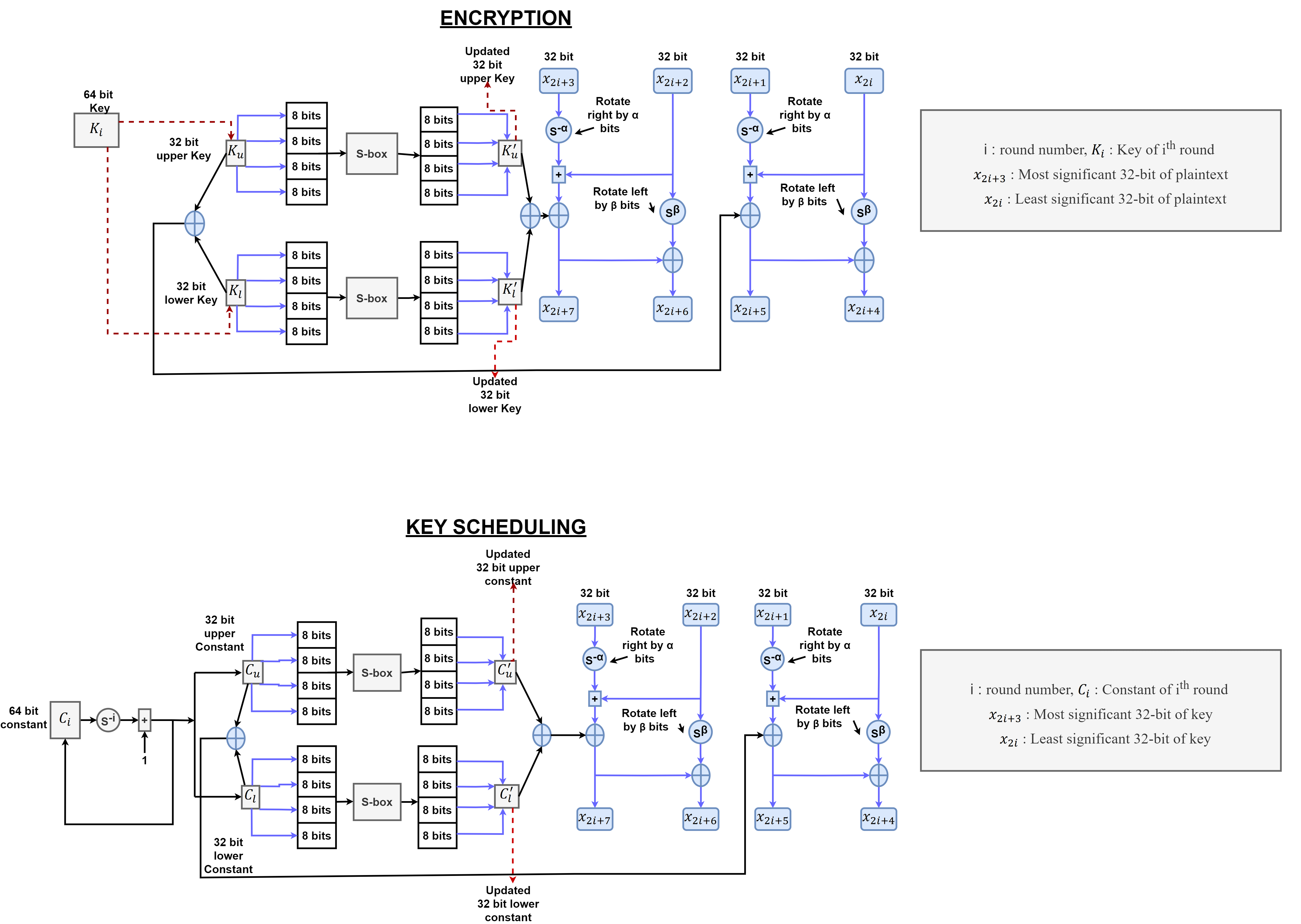}
    \caption{The figure shows the MAVShield key scheduling and encryption functions.}
    \label{D9}
\end{figure*}

\paragraph{Extraction Scheme} 
After each iteration, four 32-bit words of the pseudo-random keystream block are generated as follows: 
\begin{align}
    s_{j,i}^{[15\ldots0]} = x_{2j,i}^{[15\ldots0]} \oplus x_{(2j+5) \text{ mod 8},i}^{[31\ldots16]}, \\ 
    s_{j,i}^{[31\ldots16]} = x_{2j,i}^{[31\ldots16]} \oplus x_{(2j+3) \text{ mod 8},i}^{[15\ldots0]}. 
\end{align}
Finally, at each iteration $i$, the 128-bit ciphertext is obtained by the XOR operation of the plaintext and the keystream. The same process applies in reverse, allowing retrieval of the plaintext from the ciphertext.

\section{Proposed Cipher}\label{section5}

MAVShield is an innovative symmetric key block cipher, which we use in CTR mode with a 128-bit key and block size. 
Fig. \ref{D9} illustrates the mechanism underlying the MAVShield encryption scheme. This encryption scheme performs a dynamic process that systematically updates each 64-bit round key generated by the key-scheduling algorithm of MAVShield. The process involves a series of sequential operations, including word splitting, S-box substitution, and XOR operations. At both ends of these operations the XORed secret key acts as the input $32$-bit key for the Speck round functions. These updated $32$-bit keys are applied to each round function to transform the plaintext into the ciphertext. Also, as shown in the key scheduling part of Fig. \ref{D9}, the generation of each round key follows similar operations, followed by rotation and arithmetic operations on a 64-bit nonce. This process enhances the randomization of the secret key. 

As discussed in Section \ref{section4.2.3}, the Speck family of block ciphers is characterized by its security, flexibility, and lightweight design. It delivers outstanding performance on software platforms, ensuring the advertised (but not excessive) level of protection when resources are scarce, making it ideal for UAV-GCS MAVLink communications. MAVShield is designed by incorporating several key modifications into Speck, which improve its performance while preserving its strong security, as we now explain. 

The proposed cipher, MAVShield, is designed to enhance encryption by refining the Speck round function, which operates with a block size of 64 bits and a key size of $128$ bits. In MAVShield, the number of rounds is reduced to $10$, from $27$ rounds in Speck $64/128$. This notable reduction enhances CPU performance by accelerating the ciphering process, while ensuring that MAVShield remains secure, as shown in Section \ref{SC:security:analysis}.

MAVShield incorporates some preliminary values, including a 128-bit initialization vector (IV), a 64-bit nonce, and a 128-bit secret key. The primary objective of MAVShield is to achieve a balance between security and performance. MAVShield addresses this balance by enhancing the round number used as input to the Speck round function during key scheduling. In the original Speck round function, the round number corresponds to the total number of previous rounds. In contrast, MAVShield adds complexity by utilizing a random round number for each iteration, making it complicated  for adversaries to uncover the round keys. The encryption scheme of the proposed cipher further incorporates non-linearity via the substitution technique, which effectively obfuscates the key's impact and makes it challenging for an attacker to deduce the key from the ciphertext.

\begin{algorithm}
\caption{Pseudo-code of MAVShield Encryption}\label{alg2}
\textbf{Input:} $initial\_secret\_key \in {\{0,1\}}^{128} $, $PT$ \Comment{(PT stands for Plaintext)}\\ 
\textbf{Output: $CT$}  \Comment{(CT stands for Ciphertext)}
\begin{algorithmic}[1]
\State $K[T] \gets \ref{alg:ks}{(initial\_secret\_key)}$ \Comment{(Key schedule containing T values)}
\Function{\textnormal{$MAVShield\_xcrypt$}}{$PT, initial\_secret\_key$} \funclabel{alg:xc}
\State $\text{Initialization: } CT \gets PT$
\State $CT_3, CT_2, CT_1, CT_0 \gets \text{$128bit\_to\_32bit\_word$}(CT)$
\For{$i \gets 0$ to $(T-1)$} \Comment{(No. of rounds is $T$, which is equal to $10$)}
          \State $(K_u,K_l) \gets \ref{alg:a}{(K[i])}$
         \State $(CT_3, CT_2) \gets \ref{alg:b}{(CT_3, CT_2, K_u)}$ \Comment{(First Round Function)}
         \State $(CT_1, CT_0) \gets \ref{alg:b}{(CT_1, CT_0, K_l)}$ \Comment{(Second Round Function)}
        \State{$CT \gets (CT_{3},CT_{2},CT_{1},CT_{0}) $}
    \EndFor
    \State \Return $CT$
\EndFunction    
\end{algorithmic}
\end{algorithm}

Algorithm~\ref{alg2} outlines the pseudo-code for the MAVShield encryption process, which is divided into three main components: Key Scheduling, Round Value Generation, and the Round Function. Utilizing a 128-bit initial secret key, Algorithm \ref{alg3} generates a key schedule comprising 10 keys, each 64 bits in length. During each iteration of the encryption process, the function \ref{alg:a} is called, taking each key as input to produce an upper 32-bit word ($K_u$) and a lower 32-bit word ($K_l$).
 Subsequently,  the $Ciphertext$ array is updated  using two calls to the Speck basic round function-- one call each with the keys $K_u$ and $K_l$. The round function depends on the output of the previous round and a key generated by a key-scheduling algorithm \cite{biham1991differential}.

In  Algorithm \ref{alg3}, we carry out key scheduling by utilizing a 64-bit nonce, which undergoes rotation and complement operations. The function \ref{alg:a} refines this 64-bit word through a complex process involving word splitting, S-box substitution, and XOR operations, ultimately generating upper and lower-order words. The S-box is a non-linear substitution table that transforms a given set of input bits into a corresponding set of output bits. Within a 32-bit word, each of its four bytes acts as an index in the S-box table, where it is replaced with a corresponding 8-bit output. This ensures that all four bytes undergo individual substitution, producing four distinct output bytes. To enhance diffusion, XOR operations are applied to mix the upper and lower-order 32-bit words. The resulting XORed output words are then used as a substitute for a round number in two instances of \ref{alg:b}, enabling the generation of the key array.

\begin{algorithm}
\caption{Pseudo-code of MAVShield Key Schedule Generation}\label{alg3}
\textbf{Input: }  $initial\_secret\_key \in {\{0,1\}}^{128} $, $C_0 =  Nonce \in  {\{0,1\}^{64}}$\\
\textbf{Output: $KS[T]$ } \Comment{(Key Schedule containing T values)}
\begin{algorithmic}[1] 
\Function{\textnormal{$key\_generation$}}{$initial\_secret\_key$} \funclabel{alg:ks}
\State $\text{Initialization: }$
\State $key \gets initial\_secret\_key$
\State $key_3, key_2, key_1, key_0 \gets \text{128bit\_to\_32bit\_word}(key)$
\For{$i \gets 0$ to $(T-1)$} \Comment{(No. of rounds is $T$, which is equal to $10$)}
     \State $\text{Add \& Rotate Operations:} $
     \State $C_i \gets S^{-i}C_i$
     \State $C_i \gets complement(C_i)$
     \State $(C_u,C_l) \gets \ref{alg:a}{(C_i)}$
         \State $key_3, key_2 \gets \ref{alg:b}{(key_3, key_2, C_u)}$
         \State $key_1, key_0 \gets \ref{alg:b}{(key_1, key_0, C_l)}$
         \State $k_{i} \gets (key_{1}key_{0}) $
         \State$k_{i+1} \gets (key_{3}key_{2}) $ 
         \State $d \gets 32\_bit\_array\_to\_64\_bit\_word(k_i, k_{i+1})$
         \State $\text{Append }d \text{ to } KS[T]$
 \EndFor
        \State \Return $KS[T]$
\EndFunction
 \Function{\textnormal{RoundValueGeneration}}{$C_i$} \funclabel{alg:a}
    \Comment{($C_i$ is a 64-bit word)}
     \State $\text{Word Splitting \& S-box Substitution: } $
    \Comment{(sbox is a $256 \times 256$ matrix having each entry of 8-bit size.)}
      \State $(C_u,C_l) \gets 64\text{bit}\_\text{to}\_32\text{bit}\_\text{word}(C_i)$
     \State $(C_{u1},C_{u2},C_{u3},C_{u4}) \gets \text{32bit\_to\_8bit\_array}(C_u)$
      \State$(C_{u1},C_{u2},C_{u3},C_{u4}) \gets \text{sbox}(C_{u1},C_{u2},C_{u3},C_{u4})$
       \State $C'_u \gets \text{8bit}\_\text{array}\_\text{to}\_\text{32bit}\_\text{word}(C_{u1},C_{u2},C_{u3},C_{u4})$
        \State $(C_{l1},C_{l2},C_{l3},C_{l4}) \gets 32\text{bit}\_\text{to}\_8\text{bit}\_\text{array}(C_l)$
      \State$(C_{l1} , C_{l2}, C_{l3}, C_{l4}) \gets \text{sbox}(C_{l1} , C_{l2}, C_{l3}, C_{l4})$
       \State $C'_l \gets 8\text{bit}\_\text{array}\_\text{to}\_32\text{bit}\_\text{word}(C_{l1},C_{l2},C_{l3},C_{l4})$
        \State $\text{XOR Operations:} $
        \State$C_l \gets C_u \oplus C_l$
        \State$C_u \gets C'_u \oplus C'_l$
        \State \Return $C_u,C_l $
  \EndFunction

 \Function{\textnormal{RoundFunction}}{$a,b,key$} \funclabel{alg:b}  \Comment{($a$, $b$ $\&$ $key$ are 32-bit words)}
  \State $a \gets S^{-\alpha}a$ \Comment{($a$ is right shifted by $\alpha$)}
  \State $a \gets a + b$
  \State $a \gets a \oplus key$
  \State $b \gets S^{\beta}b$ \Comment{($b$ is left shifted by $\beta$)}
  \State $b \gets b \oplus a$
  \State \Return $a,b$
  \EndFunction
\end{algorithmic}
\end{algorithm}

This approach ensures a high degree of randomness in key scheduling, such that even a minor change in the key leads to a significant and unpredictable alteration in the ciphertext during encryption. In other words, these operations enhance the encryption's confusion and diffusion properties \cite{perlman2016network}, thereby improving its security.

\section{Security Analysis} 
\label{SC:security:analysis}

In this section, the security of MAVShield is analyzed using differential cryptanalysis  \cite{biham1991differential} as well as using the Wireshark network packet analyzer \cite{sanders2017practical}. 

\subsection{Differential Cryptanalysis}
The goal of cryptanalysis is to find weaknesses in encryption algorithms, which could allow an attacker to decrypt messages or gain unauthorized access to protected information. The encryption algorithm is assumed to be publicly known, and the data security follows solely from the secrecy of the randomly selected key. 

Differential cryptanalysis, first introduced in \cite{biham1991differential}, has become a powerful technique for successfully attacking a wide array of block ciphers. In a block cipher, data is processed in fixed-sized blocks, with the encryption algorithm using a secret key to convert each block of $Plaintext$ into $Ciphertext$. In its basic form, differential cryptanalysis examines the effect of particular differences in $Plaintext$ pairs on the differences of the resultant $Ciphertext$ pairs. These attacks are statistical, where the attacker seeks to identify probabilistic patterns over multiple rounds of the cipher. By distinguishing the cipher's behavior from a random permutation, the attacker aims to recover the secret key.

In the model in  this paper, the adversary attempts to infer the confidential secret key ($K$) used to encrypt the $Plaintext$ pair ($P$ and $P'$), which, after the encryption process, is transformed into the $Ciphertext$ pair ($C$ and $C'$). This method generally relies on analyzing many pairs of $Plaintext$ with a consistent specific difference, using only the corresponding $Ciphertext$ pairs for the analysis.

In differential cryptanalysis, the attacker tries to identify differential characteristics ($\Delta P$, $\Delta C$) such that the difference between $C$ and $C'$ is linked to the difference between $P$ and $P'$: 
\begin{equation*}
    \underbrace{P \oplus P'}_{\text{Difference}\text{ in Plaintexts}} = \hspace{1mm} \Delta P \hspace{3mm} \longrightarrow \hspace{3mm} \Delta C \hspace{1mm} =  \underbrace{C \oplus C'}_{\text{Difference}\text{ in Ciphertexts}} 
\end{equation*} 
The differential characteristics are valid if, for many $Plaintext$ pairs for which the difference is $\Delta P$, the resulting ciphertext pairs frequently exhibit the difference $\Delta C$. By analyzing the pairs that produce the output difference, the attacker might succeed in identifying characteristics that distinguish the cipher \cite{albrecht2009algebraic}.

\subsubsection{Chosen Plaintext Attack (CPA) Model}
 Differential cryptanalysis is performed using the $CPA$ model to analyze the security of MAVShield. In this model, the attacker has the ability to choose arbitrary $Plaintexts$ and obtain their corresponding $Ciphertexts$ from the encryption system. For this, we make the following assumptions \cite{sabuwala2024approach}:
 
\begin{itemize}
    \item The adversary has access to the proposed algorithm,  which he/ she uses to perform encryption.
    \item $Plaintext$ pairs ($P_i$, $P_i'$) are chosen to differ by only one bit, where $i$ is the bit position index.
    \item The set of $Plaintext$-$Ciphertext$ pairs available to the attacker are denoted by: ($P_i$, $C_i$) and ($P'_i$, $C_i'$), where $P_i$, $P_i'$ represent differential $Plaintexts$ and $C_i$, $C_i'$ are the corresponding encrypted outputs.
    \item The attacker has no prior knowledge of the secret key employed by the algorithm.
\end{itemize}
\vspace{1mm}

\begin{algorithm}
\caption{Pseudo-code for generating $Plaintext$ Pairs with Unit Distance:}\label{alg5}
\textbf{Output: $\text{`pt.bin'}$} \Comment{(File containing plaintext pairs of unit distance)}
\begin{algorithmic}[1]
\For{$q \gets 0$ to $3,000,000-1$}
    \State $P[q] \gets \ref{alg:a1}(\text{ })$
    \State $P'[q] \gets P[q]$
    \State $P'[q] \gets \ref{alg:a2}(P'[q])$
    \State $\text{`pt.bin'} \gets (P[q],P'[q])$ \Comment{(Plaintext pairs are appended to a binary file)}
\EndFor
\State \Return $\text{`pt.bin'}$
\Function{\textnormal{GenerateRandomPlaintexts}}{ } \Comment{(Generate a random 32-byte array of plaintext)} \funclabel{alg:a1}  
    \For{$i \gets 0$ to $31$}
        \State $PT[i] \gets \text{Generate a random byte between 0 to 255}$
        \State $PT[i] \gets \ref{alg:a3}(PT[i])$ 
    \EndFor
    \State \Return $PT$
\EndFunction
\Function{\textnormal{FlipRandomBit}}{$B$} \Comment{(Single bit of 32-byte input array is inverted)} \funclabel{alg:a2}  
    \State $i \gets \text{random integer between 0 to 31}$
    \State $B[i] \gets (B[i] \oplus 1)$ \Comment{(least significant bit of selected byte is flipped)}
    \State \Return $B$
\EndFunction
\Function{\textnormal{ByteToAsciiBit}}{$byte$} \funclabel{alg:a3} \Comment{(An unsigned integer byte is converted into ASCII representation)}
    \For{$i \gets 7$ to $0$}
        \State $byte \gets ((byte \gg i) \oplus 1)$
    \EndFor
    \State \Return $byte$
\EndFunction
\end{algorithmic}
\end{algorithm}

\subsubsection{Unit Distance Plaintext Pairs}

Algorithm \ref{alg5} describes the procedure for generating 32-byte $Plaintext$ pairs that differ by a single bit. This is accomplished by inverting a single bit through a XOR operation with $1$. The $Plaintext$ pairs can either be randomly generated or assumed to be selected by an adversary, who then obtains the corresponding $Ciphertext$ by encrypting the $Plaintext$ using the MAVShield algorithm along with a randomly selected confidential secret key. The adversary aims to intercept the $Ciphertext$ pairs to recover the secret key.

\subsection{Analysis using NIST and Diehard Statistical Suites}

 NIST \cite{rukhin2001statistical} and Diehard \cite{bogos2022remark} are both test suites, which are a collection of statistical tests used to evaluate the randomness of binary sequences produced by cryptographic random number generators. To carry out this analysis, the following approach has been used:
 \begin{enumerate}[-]
     \item We produce 3,000,000 pairs of $Plaintext$ that exhibit minimal differences in terms of bits.
     \item We apply the proposed algorithm to produce 3,000,000 pairs of $Ciphertext$ for the corresponding $Plaintext$ pairs.
     \item To determine if statistical differences exist, $Ciphertext$ pairs are thoroughly analyzed using these test suites.
     \item If statistical differences are observed in the ciphertext produced by the proposed algorithm, it indicates a flaw or weakness in the encryption process. Therefore, there should be no discernible patterns in the encrypted output.
 \end{enumerate}

 \begin{algorithm}
\caption{Pseudo-code for generating $Ciphertext$ Pairs:}\label{alg6}
\textbf{Input: $(P,P') \in \text{`pt.bin'}$, $secret\_key \in {\{0,1\}}^{128}$}\\
\textbf{Output: $\text{`ct.bin'}$} \Comment{(File containing ciphertext pairs)}
\begin{algorithmic}[1]
\For{$q \gets 0$ to $3,000,000-1$}
    \State $C[q] \gets \textbf{$MAVShield\_xcrypt(P[q], secret\_key)$}$
    \State $C'[q] \gets \textbf{$MAVShield\_xcrypt(P'[q], secret\_key)$}$
    \State $\text{`ct.bin'} \gets (C[q],C'[q])$ \Comment{(Ciphertext pairs are appended to a binary file)}    
 \EndFor
 \State \Return $\text{`ct.bin'}$
\end{algorithmic}
\end{algorithm}

Algorithm \ref{alg6} shows the procedure for generating $Ciphertext$ pairs, which are written to the binary file ``ct.bin''. The $Ciphertext$ pairs were subjected to both the NIST Statistical Test Suite (NSTS) and the Diehard battery of statistical tests. The evaluation focused on $p$-values and randomness metrics. For NIST, the minimum pass rate for each statistical test is $8$ for a sample size of $10$ binary sequences, with $p$-values needing to exceed $0.025$ to be considered successful. In contrast, the Diehard tests require $p$-values to fall within the interval of $0.025$ to $0.975$ for a successful outcome \cite{pervushin2021quantum}.

\begin{table}[!ht]
\caption{The table shows the NIST test results.}
\label{table:II}
\centering
\begin{tabular}{|l|l|l|}
\hline
\rowcolor[HTML]{9698ED} 
\textbf{Tests} & \textbf{$p$-value} & \textbf{Proportion} \\ \hline
\rowcolor[HTML]{FFFFFF} 
\textbf{Frequency} & \textbf{0.350485} & \textbf{10/10} \\ \hline
\rowcolor[HTML]{FFFFFF} 
\textbf{Block Frequency} & \textbf{0.350485} & \textbf{10/10} \\ \hline
\rowcolor[HTML]{FFFFFF} 
\textbf{Cumulative Sums} & \textbf{0.739918} & \textbf{10/10} \\ \hline
\rowcolor[HTML]{FFFFFF} 
\textbf{Runs} & \textbf{0.350485} & \textbf{10/10} \\ \hline
\rowcolor[HTML]{FFFFFF} 
\textbf{Longest Runs} & \textbf{0.035174} & \textbf{10/10} \\ \hline
\rowcolor[HTML]{FFFFFF} 
\textbf{Linear complexity} & \textbf{0.911413} & \textbf{10/10} \\ \hline
\rowcolor[HTML]{FFFFFF} 
\textbf{Approximate Entropy} & \textbf{0.213309} & \textbf{10/10} \\ \hline
\rowcolor[HTML]{FFFFFF} 
\textbf{Overlapping template} & \textbf{0.911413} & \textbf{9/10} \\ \hline
\rowcolor[HTML]{FFFFFF} 
\textbf{Non overlapping template} & \textbf{0.991468} & \textbf{10/10} \\ \hline
\rowcolor[HTML]{FFFFFF} 
\textbf{Serial} & \textbf{0.991468} & \textbf{10/10} \\ \hline
\end{tabular}
\end{table}

Table \ref{table:II} presents the results of $10$ statistical tests, which are part of the NIST test suite, illustrating the performance of the MAVShield encryption algorithm. These tests assess whether the output of the cryptographic function aligns with the statistical properties of a truly random process \cite{rukhin2001statistical}. 

Among the tests conducted, the frequency test evaluates the proportion of zeros and ones with a $p$-value of $0.350485$ and with a maximum pass rate of $10$, and the frequency within a block test refines this approach by assessing the proportion of ones within fixed-length blocks, yielding a $p$-value of $0.350485$ \cite{freqtest}. The cumulative sum test, which interprets the sequence as a random walk and examines how closely its excursions remain near zero, produces a $p$-value of $0.739918$. The runs test analyzes ``runs" of consecutive $1$s and $0$s to determine if their oscillation is excessively fast or slow. In contrast, the longest runs test verifies whether the longest sequence of $1$s aligns with expectations for randomness. These two tests yield $p$-values of $0.350485$ and $0.035174$, respectively.

The linear complexity test determines whether or not the sequence is complex enough to be considered random, while the approximate entropy test compares the frequency of overlapping blocks of two consecutive lengths against the expected result for a random sequence; these tests produce $p$-values of $0.911413$ and $0.213309$, respectively. 

Similarly, the overlapping template test rejects sequences that show too many or too few occurrences of runs of $1$s, generating a $p$-value of $0.911413$. The non-overlapping template test, which evaluates the occurrences of predefined target strings \cite{li2020parallel}, produced a high $p$-value of $0.991468$. Likewise, the serial test determines whether the number of occurrences of the $2^m$ $m$-bit overlapping patterns is approximately the same as expected for a random sequence \cite{bogos2022remark}; it returns a $p$-value of $0.991468$, suggesting weak statistical evidence for any observed differences. 
The analysis shows that all 3,000,000 $Ciphertext$ pairs meet the criteria of the tests, demonstrating that MAVShield is secure.

\begin{table}[!ht]
\caption{The table shows the Diehard test results.}
\label{table:VI}
\centering
\begin{tabular}{|l|l|l|}
\hline
\rowcolor[HTML]{9698ED} 
\textbf{Tests} & \textbf{$p$-value} & \textbf{Assessment} \\ \hline
\rowcolor[HTML]{FFFFFF} 
\textbf{Birthday Spacing Test} & \textbf{0.724920} & \textbf{Passed} \\ \hline
\rowcolor[HTML]{FFFFFF} 
\textbf{OPERM5} & \textbf{0.849769} & \textbf{Passed} \\ \hline
\rowcolor[HTML]{FFFFFF} 
\textbf{Binary Rank Test ($32 \times 32$)} & \textbf{0.054433} & \textbf{Passed} \\ \hline
\rowcolor[HTML]{FFFFFF} 
\textbf{Binary Rank Test ($6 \times 8$)} & \textbf{0.725270} & \textbf{Passed} \\ \hline
\rowcolor[HTML]{FFFFFF} 
\textbf{OPSO} & \textbf{0.194492} & \textbf{Passed} \\ \hline
\rowcolor[HTML]{FFFFFF} 
\textbf{OQSO} & \textbf{0.194211} & \textbf{Passed} \\ \hline
\rowcolor[HTML]{FFFFFF} 
\textbf{DNA} & \textbf{0.180828} & \textbf{Passed} \\ \hline
\rowcolor[HTML]{FFFFFF} 
\textbf{Bitstream Test} & \textbf{0.253896} & \textbf{Passed} \\ \hline
\rowcolor[HTML]{FFFFFF} 
\textbf{Count-The-1's Test on a stream of bytes} & \textbf{0.459584} & \textbf{Passed} \\ \hline
\rowcolor[HTML]{FFFFFF} 
\textbf{Count-The-1's Test for specific byte} & \textbf{0.626511} & \textbf{Passed} \\ \hline
\textbf{Parking Lot Test} & \textbf{0.069298} & \textbf{Passed} \\ \hline
\rowcolor[HTML]{FFFFFF} 
\textbf{3D Sphere Test} & \textbf{0.084122} & \textbf{Passed} \\ \hline
\rowcolor[HTML]{FFFFFF} 
\textbf{Squeeze Test} & \textbf{0.650560} & \textbf{Passed} \\ \hline
\rowcolor[HTML]{FFFFFF} 
\textbf{Craps Test} & \textbf{0.597665} & \textbf{Passed} \\ \hline
\end{tabular}
\end{table}

Next, in our security analysis of the MAVShield cipher, the Diehard test suite is employed to rigorously assess the quality of randomness through 14 statistical evaluations. The results are shown in Table \ref{table:VI}. The birthday spacing test analyzes the distribution of gaps between $m$ random birthdays in a year of $n$ days, comparing it to the expected distribution to assess the randomness quality, giving a $p$-value of $0.724920$ \cite{diehard}. The overlapping 5-permutation test (OPERM5) analyzes five-number sequences in random integers using covariance-based statistics. Its $p$-value of $0.849769$ indicates that there is no significant deviation from the expected random distribution. Binary rank tests analyze matrix ranks by treating columns as axes in an $N$-dimensional cube. Here, the chi-squared test results in $p$-values of $0.054433$ (for $32\times32$ matrices) and $0.725270$ (for $6\times8$ matrices) by comparing the observed and expected rank distributions. These values indicate that deviations are not statistically significant.

In Diehard, the test groups Bitstream, Overlapping Pairs Sparse Occupancy (OPSO), Overlapping Quadruples Sparse Occupancy (OQSO), and DNA are very similar. The OPSO, OQSO, and DNA tests assess randomness by counting missing words in generated sequences. The OPSO test (2-letter words, 10-bit segments) yields a $p$-value of $0.194492$  over 23 runs, shifting bits from 1–10 to 23–32 across $2^{21}$+$1$ keystrokes. The OQSO test (4-letter words, 5-bit segments) produces a $p$-value of $0.194211$  over 28 runs, shifting from 1–5 to 28–32 across $2^{21}$+$3$ keystrokes. DNA (10-letter words, 2-bit segments) yields a $p$-value of $0.180828$ by analyzing 4 letters {C, G, A, T}. The Bitstream test treats the file as a stream of bits, forming overlapping 20-letter words. Missing words are counted in $2^{21}$ overlapping sequences, resulting in a $p$-value of $0.253896$.

 The Count the 1's tests map bytes to letters (A–E) based on their $1$s count. This test on a stream of bytes analyzes $5^5$ overlapping words in 256,000 sequences, yielding a $p$-value of $0.459584$, while the test on specific bytes shifts bytes across 25 runs (bits 1–8 to 25–32), producing a $p$-value of $0.626511$. These $p$-values indicate that the observed distributions align well with the expected randomness, showing no significant deviation from uniformity.

The Parking Lot test simulates parking a unit-radius car in a $100 \times 100$ square. Each car is randomly placed, and if a collision occurs, a new position is chosen. After 12,000 attempts, the number of successfully parked cars follows a normal distribution and generates a $p$-value of $0.069298$. The 3D Sphere test places $4000$ points in a $10000^3$ cube, forming spheres to reach the nearest point. The smallest sphere’s volume follows an exponential distribution. Transforming the minimum cube radius maps values to a uniform  $[0,1)$ distribution, validated by a  Kolmogorov-Smirnov (K-S) test, yielding a $p$-value of $0.084122$. 

The Squeeze test progressively ``squeezes'' the initial value $k = 2^{32}$  down to zero by repeatedly multiplying it by a uniform random number $U$ in $[0,1)$ until $k=0$, analyzing the iteration counts $J$ with a chi-square test, and producing a $p$-value of $0.650560$. The Craps test simulates 200,000 games, counting wins and throws per game. Wins follow a normal distribution, while throws undergo a chi-square test. Dice rolls derive from floating 32-bit integers. The test yields a $p$-value of $0.597665$. $p$-values near $0$ or $1$ suggest possible non-randomness, and extreme cases ($p<0.0025$ or $p>0.9975$) indicate test failure at the $0.05$ significance level. Since this is not the case, MAVShield has successfully passed all the Diehard tests, affirming its robustness and reliability in ensuring cryptographic security.

\subsection{Security Analysis using Wireshark} 

Wireshark \cite{sanders2017practical} is a network packet analyzer that  runs on the same laptop as the GCS in our testbed. Wireshark captures network packets in real time, allowing users to see all the data being transmitted over a network. It supports various protocols, enabling us to analyze the details of network communications. Wireshark displays packets in a structured format, showing details such as timestamps, source and destination IP addresses, protocols, and payload data \cite{sanders2017practical}. To introduce the MAVLink protocol in the Wireshark interface, a $mavlink\_2\_common.lua$ script is used as a plugin in Wireshark, enabling the parsing of MAVLink messages.

\begin{figure}[h]
    \centering
     \includegraphics[width=0.9\linewidth]{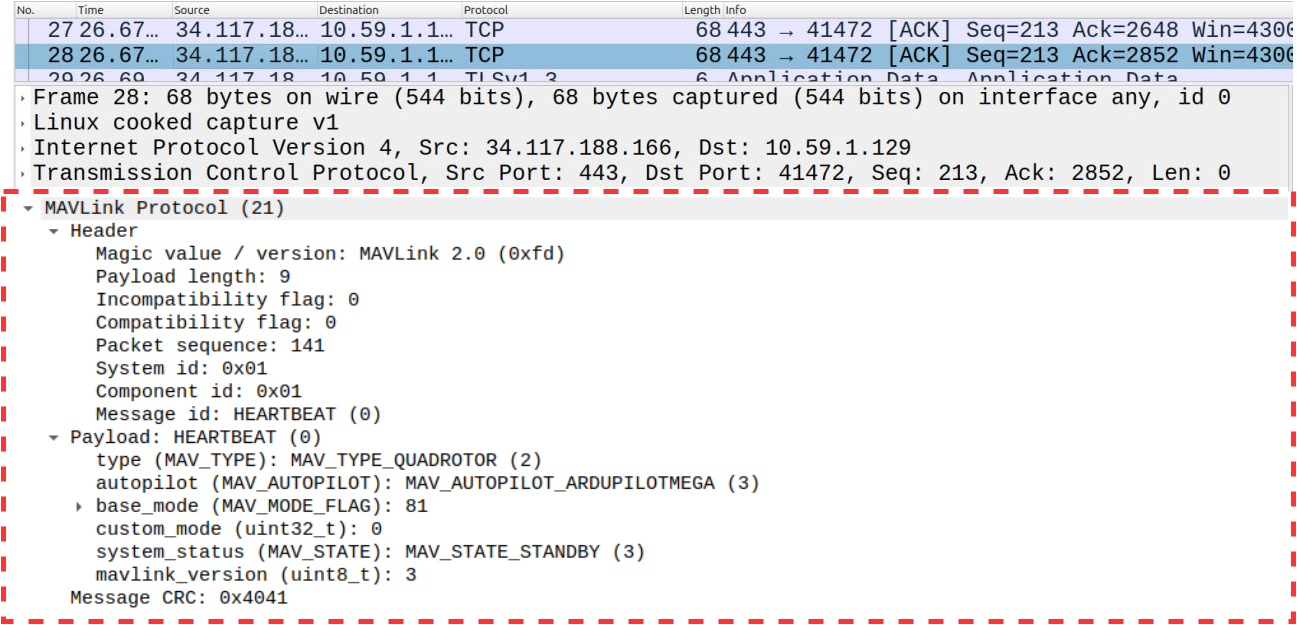}
    \caption{The figure shows a captured MAVLink 2.0  HEARTBEAT message. }
    \label{D15}
\end{figure}

\begin{figure}[h]
    \centering
     \includegraphics[width=0.9\linewidth]{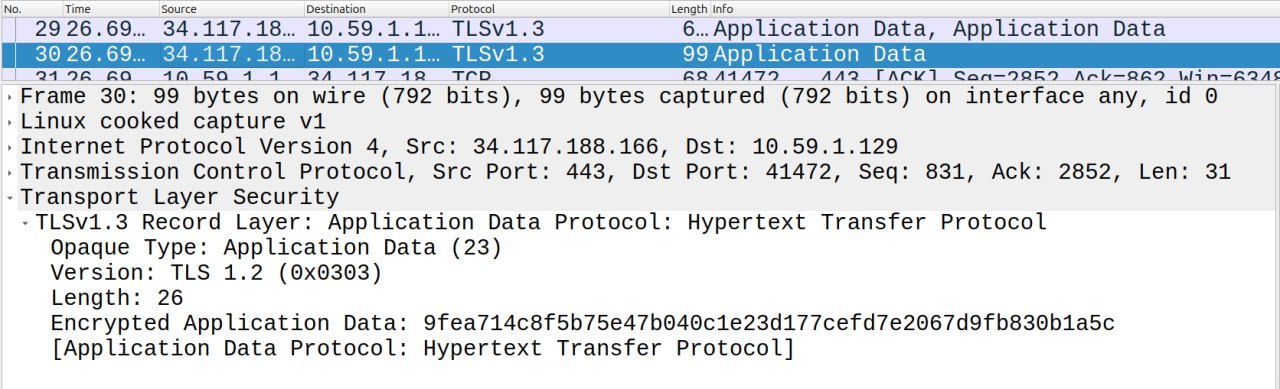}
    \caption{The figure shows a captured MAVLink 2.0  encrypted message. }
    \label{D19}
\end{figure}

 Since the transport layer encrypts all MAVLink messages during transmission if  Transport Layer Security (TLS) is used, the payload data in network packets will appear encrypted in Wireshark. We implemented the proposed encryption scheme, MAVShield, in the MAVLink protocol for encrypting payload data across all message types, except the HEARTBEAT message (message id. = 0). 
 
 Fig. \ref{D15} shows communication between the source IP address 34.117.188.166 and the destination IP address 10.59.1.129, in the case where TLS is absent, resulting in the payload data being unencrypted and visible. In contrast, Fig. \ref{D19} shows that communication is still taking place between the same source and destination addresses, but in this case, TLS is present and the application data is encrypted, obscuring the payload as ciphertext. This demonstrates the effectiveness of the proposed encryption algorithm in safeguarding sensitive information in the communication process.

\section{Technologies Used and Experimental Setup}
In this section, we describe the software and hardware technologies used and the experimental setup for our drone testbed.

\subsection{Technologies Used}
In our drone testbed, we use the following technologies:
 \begin{itemize}
     \item {\bf QGroundControl (QGC) Ground Station:}  It is an open source, C++ based GCS program, which fully supports MAVLink communication \cite{knuthwebsite} and runs on an Ubuntu 22.04 laptop.
     \item {\bf Autopilot:} It is an autonomous program that allows UAVs to perform missions with or without the intervention of a pilot. Autopilot is embedded into the UAV and allows the control of the UAV movements. We use the ArduPilot software for autopilot functions \cite{ardu}.
     \item {\bf SITL:} It is a build of autopilot code that uses the MAVLink protocol. SITL allows the simulation of a plane, a copter, and a rover without the requirement of any specific hardware. It is deployed with a simulated (virtual) UAV, and is used to verify MAVLink communication before experimentation using our hardware testbed.
 \end{itemize}
 
\subsection{Experimental Setup}

\begin{figure}[h]
    \centering
     \includegraphics[width=0.85\linewidth]{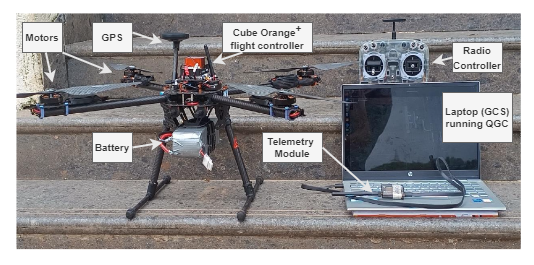}
    \caption{The figure shows the experimental setup of our drone testbed. \iffalse{\bf (DONE: Change ``Laptop (GCS) with running QGC'' to ``Laptop (GCS) running QGC''.)}\fi}
    \label{D10}
\end{figure}

Our experimental drone setup is comprised of a Quadcopter with a Pixhawk Cube Orange$^{+}$ flight controller \cite{cube}, Radiomaster boxer radio controller \cite{RC}, Holybro-SiK 433 MHz, 500 mW telemetry radio module \cite{telemetry}, Ublox GPS with antenna \cite{GPS}, LiPo 6s 10000 mAh battery \cite{LiPo}, four motors with electronic speed control \cite{motor} and a laptop running QGC as the GCS. Fig. \ref{D10} shows the main components of our testbed.

The Cube Orange$^{+}$ flight controller used in the testing environment has the following characteristics:
\begin{itemize}
    \item 
    CPU: 32-bit,  480 MHz ARM Cortex M7 processor  and 240 MHz Cortex M4 processor,
    \item 
    RAM: 1 MB, 
    \item 
    Flash: 2 MB.
\end{itemize}

\begin{figure}[h]
    \centering
     \includegraphics[width=0.9\linewidth]{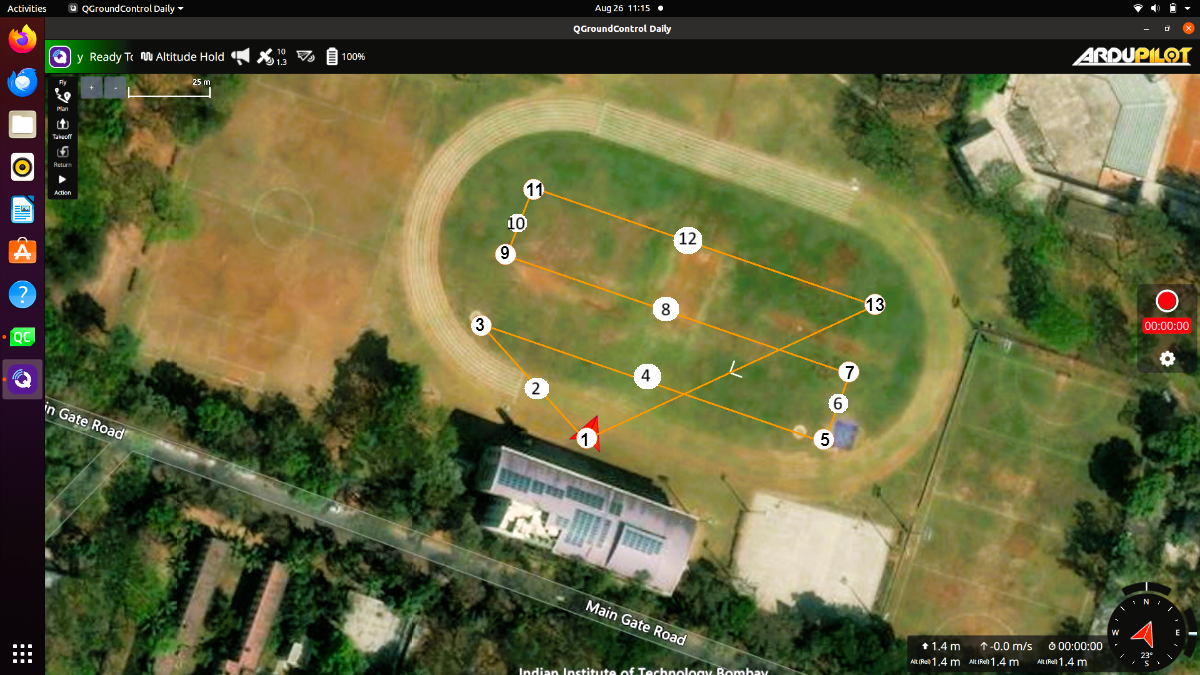}
    \caption{The figure shows the mission plan displayed in  QGC.}
    \label{D11}
\end{figure}

We outlined a mission using $13$ waypoints, as illustrated in Fig. \ref{D11}. The drone is set to take off from waypoint 1 (target location) and, upon completing the designated mission path, returns to its launch location. This mission is used as a test for all the encryption algorithms that we consider.

\section{Implementation}
In this section, we describe our drone testbed implementation. 

\subsection{Implementation of the Encryption-Decryption Process}
Since all the encryption algorithms we consider are based on symmetric key cryptography, we use the same secret key for encryption and decryption.  To exchange data securely between the GCS and the UAV, we modify MAVLink Libraries both in QGC and ArduPilot. Specifically, we redefine the $mavlink\_helpers.h$ file, which handles the transmission, encoding, decoding, and reception of messages. Our encryption and decryption schemes are incorporated in this file, and the secret key is hard-coded in the source code.  We  modify three functions in the MAVLink implementation \cite{pizzolante2023improving}:
 \begin{enumerate}
                   \item $\bf mavlink\_finalize\_message\_buffer$:
                    In this function, first encryption is performed, and then the checksum of the payload is calculated at the GCS (see Fig. \ref{fig_encryption}).  
                   \item $ \bf \_mav\_finalize\_message\_chan\_send$: This function is similar to  the $mavlink\_finalize\_message\_buffer$ function, but it is used in the UAV.    
                   \item $\bf  mavlink\_frame\_char\_buffer$: This function parses the received message, i.e., it enables the decryption of the payload message after successful checksum verification (see Fig. \ref{fig_decryption}).
                   \end{enumerate}

Also, the HEARTBEAT message (message id. = 0) transmitted from the UAV to the GCS is intentionally left unencrypted to ensure a continuous connection and to allow for real-time monitoring without compromising responsiveness.

\subsection{Mission Execution}
During the testing phase, SITL and QGC were run on the laptop. We run the $sim\_vehicle.py$ file, which starts autopilot SITL and establishes MAVLink communication between a fictitious UAV and QGC. The encrypted reading is obtained in the Ubuntu terminal while SITL runs on the same laptop.

After successfully verifying MAVLink communication in SITL, a custom firmware file, $arducopter.apj$, is generated for Cube Orange$^{+}$ using the ArduPilot program. The mission is conducted under moderate wind conditions, with the drone flying at an altitude of $25$ meters. During flight, the telemetry radio module facilitates communication between the UAV and the GCS via an RF data link operating at either 2.4 GHz or 5.8 GHz. Simultaneously, the GPS module receives signals from multiple satellites through designated RF frequencies, ensuring precise positioning and effective flight control.

\begin{figure}[h]
    \centering
     \includegraphics[width=0.98\linewidth]{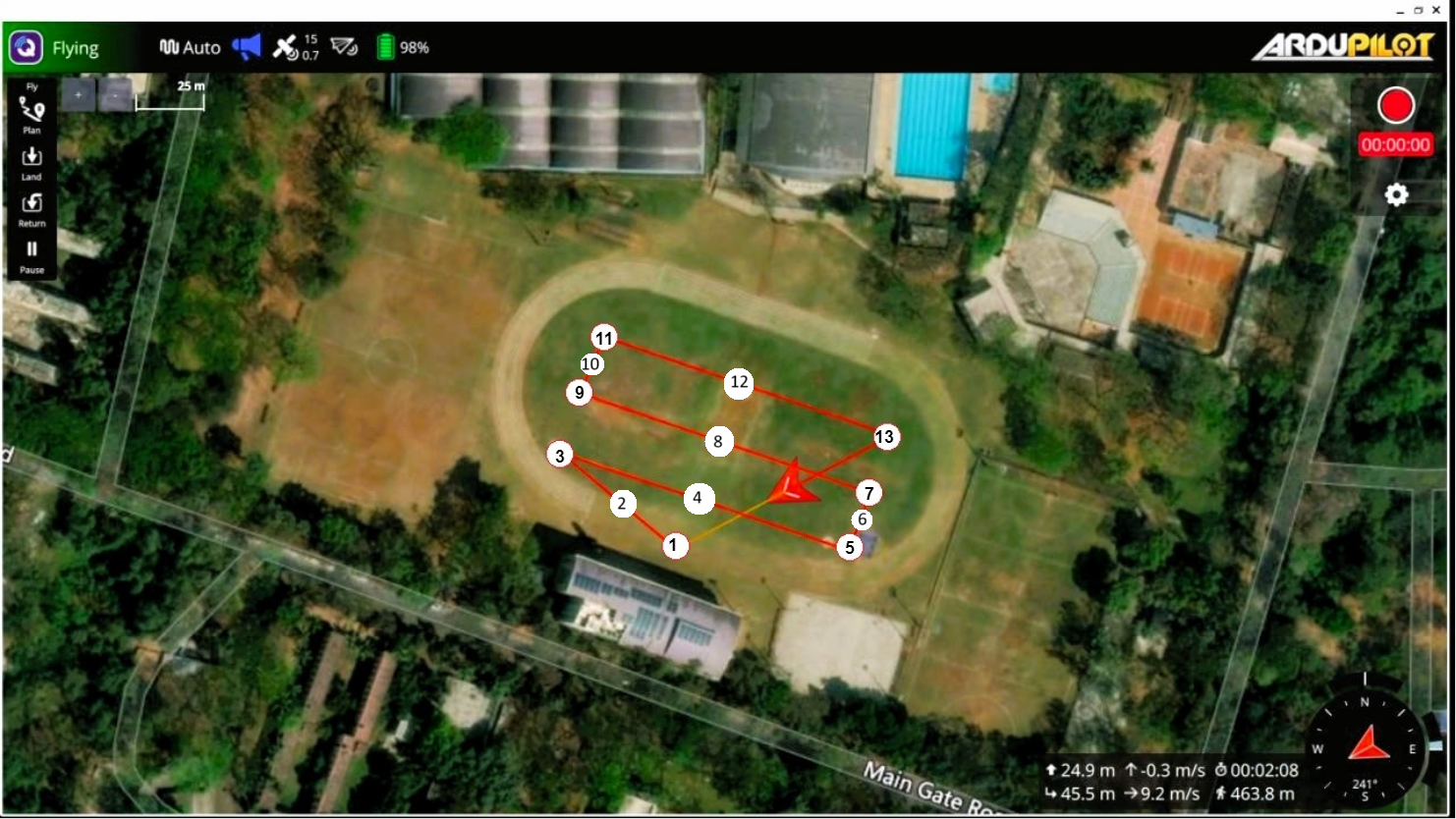}
    \caption{The figure shows the traversed mission path in QGC.}
    \label{D13}
\end{figure}

\begin{figure}[h]
    \centering
     \includegraphics[width=0.9\linewidth]{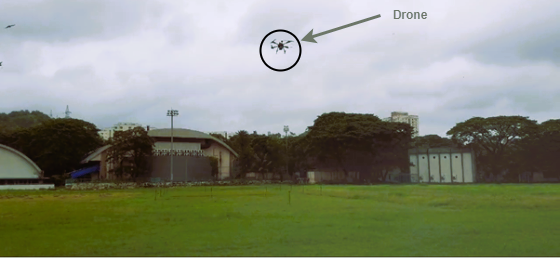}
    \caption{The figure shows the drone during the mission execution.}
    \label{D14}
\end{figure}

 Fig. \ref{D13} shows the followed mission path  in red. Also, Fig. \ref{D14} displays the actual drone flying in this path.

\section{Performance Evaluation}
The ArduPilot firmware on the drone was modified twice for ground testing: first, the standard version was used with no encryption, and then it was replaced with a version incorporating encryption. With both the versions, the same mission plan-- shown in Fig. \ref{D11}-- was executed separately in a playground in the IIT Bombay campus.

We evaluated the performance of four encryption algorithms: AES, ChaCha20, Speck, and Rabbit. In our experiments, AES and Speck operated in CTR mode with a 128-bit block size. We compared the performance of these algorithms with that of MAVShield and the unencrypted version of the MAVLink protocol. The evaluation was based on the following metrics: available memory (in bytes), CPU usage (as a percentage), and battery power consumption (in milliwatts).

For each encryption algorithm integrated into the MAVLink protocol, we conducted the same mission twice, using a real drone to evaluate its performance. We monitored the CPU usage and the available memory during each test at millisecond intervals, while also estimating the UAV's battery power consumption during each flight session. The collected data was analyzed and visualized through comparative graphs to highlight the obtained results.
 
\begin{figure}[htbp]
    \centering
     \includegraphics[width=0.9\linewidth]{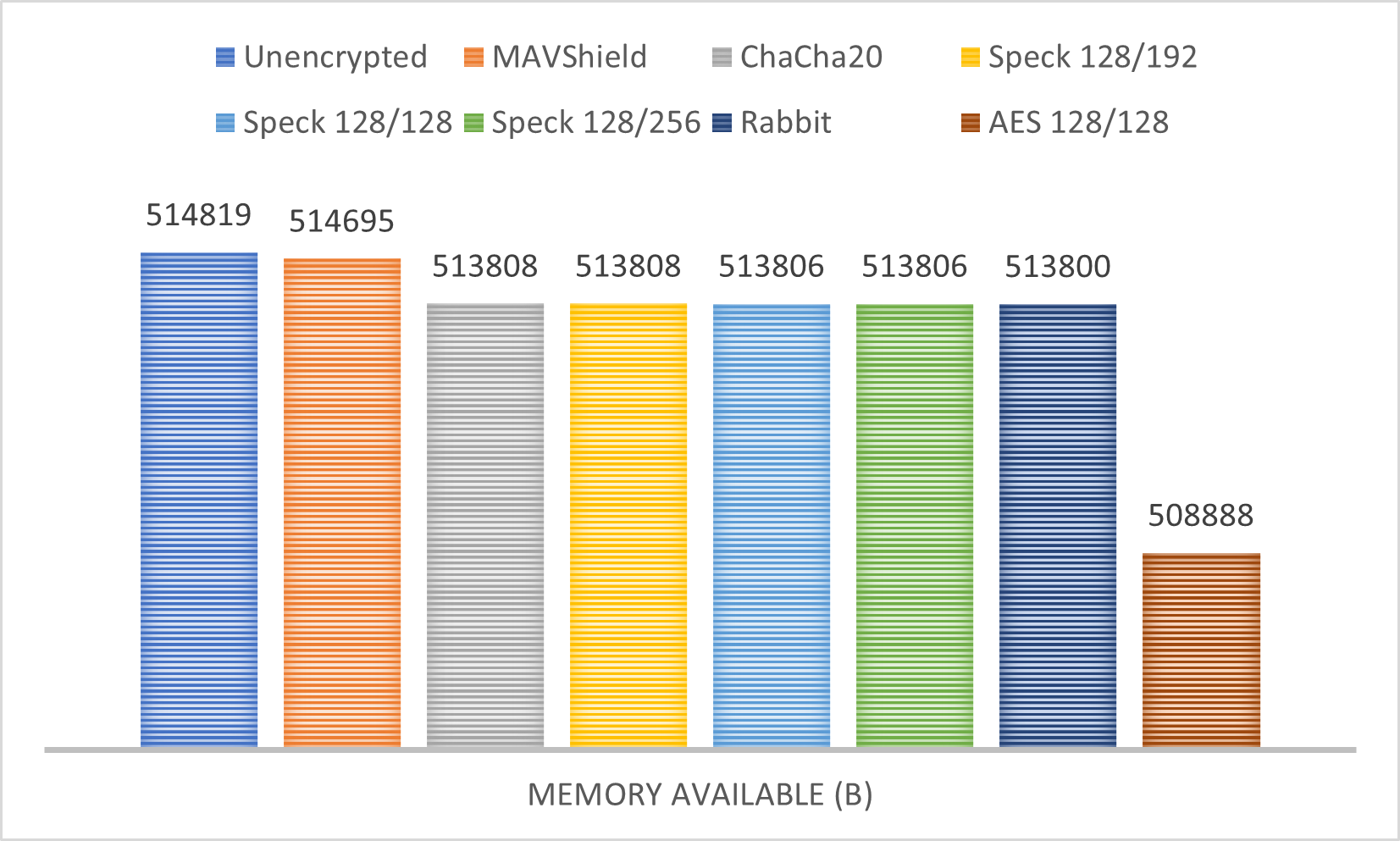}
    \caption{The figure shows the available RAM under different encryption algorithms.}
    \label{D20}
\end{figure}

\begin{figure}[htbp]
    \centering
     \includegraphics[width=0.9\linewidth]{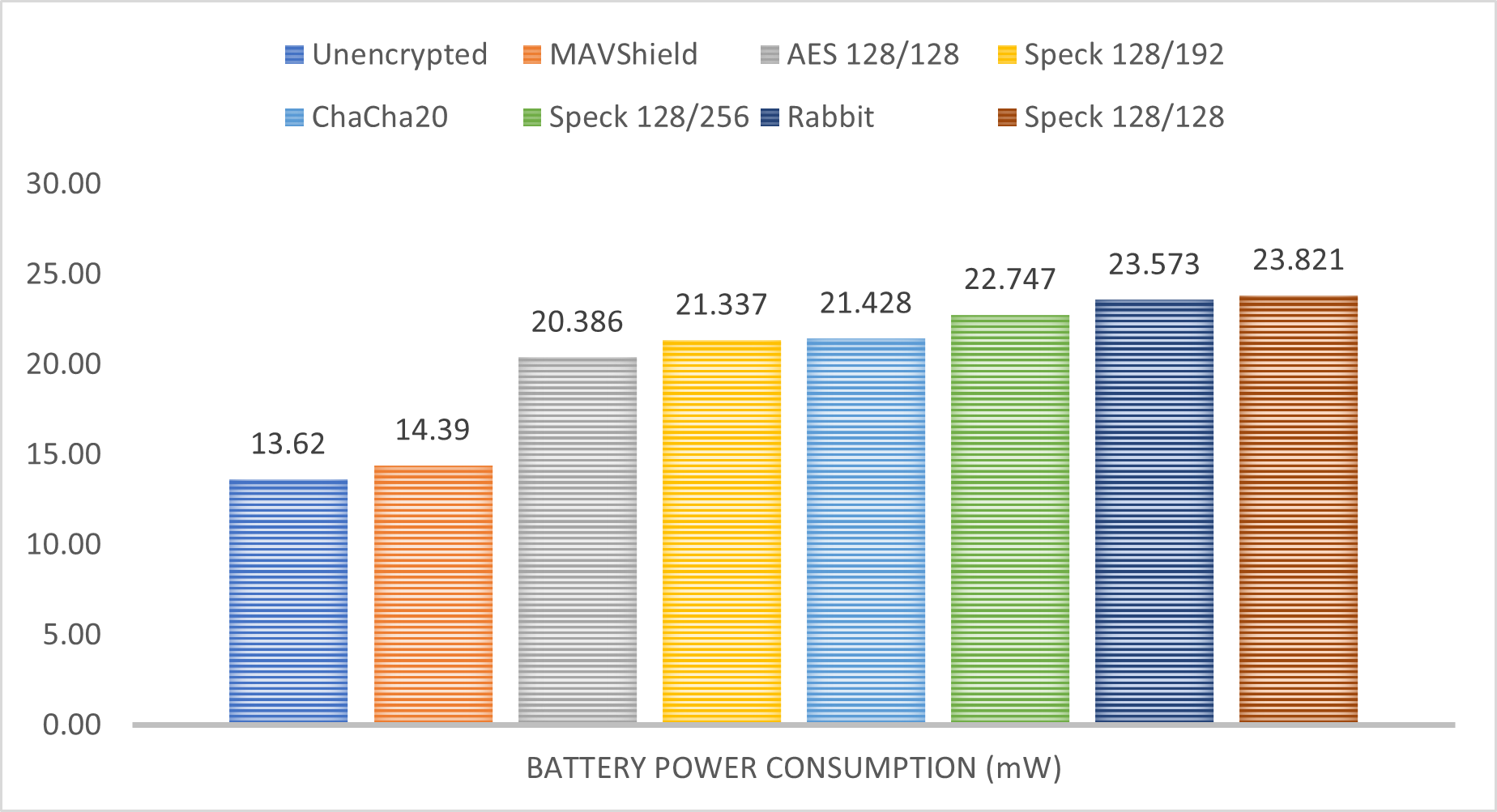}
    \caption{The figure shows the battery power consumption under different encryption algorithms.}
    \label{D21}
\end{figure}

\begin{figure}[htbp]
    \centering
     \includegraphics[width=0.9\linewidth]{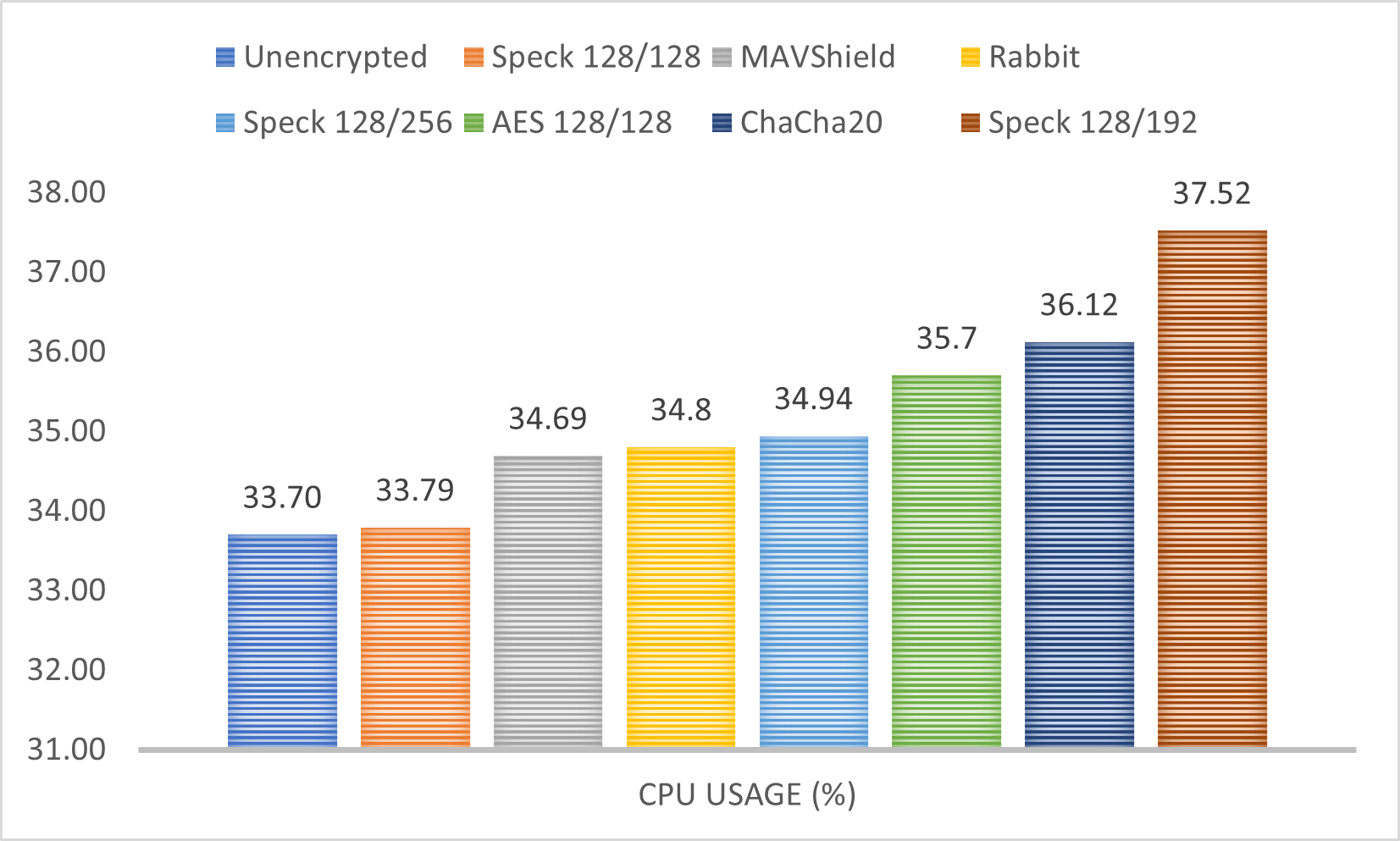}
    \caption{The figure shows the CPU usage under different encryption algorithms.}
    \label{D22}
\end{figure}

Figs. \ref{D20}, \ref{D21}, and \ref{D22} present our results for memory availability, battery power consumption, and CPU usage, respectively.  Fig. \ref{D20} depicts the available RAM in the Cube Orange$^{+}$ flight controller for different security algorithms when MAVLink communication is encrypted. The figure shows that MAVShield exhibits the lowest memory consumption among all the evaluated encryption algorithms.  Fig. \ref{D21} shows that the proposed encryption technique, MAVShield, requires only $14.39$ mW, making it the most energy-efficient option. This efficiency in power usage allows the drone to achieve a long flight duration while maintaining secure communication. 

\FloatBarrier

Fig. \ref{D22} shows that MAVShield has a CPU utilization of $34.69 \,\%$, which indicates that it has a low computational overhead and allows a lot of CPU resources for other critical UAV tasks. Overall, the bar charts clearly demonstrate that MAVShield outperforms all the other encryption algorithms. 

To enable a performance comparison, Table \ref{table:III} shows the experimental results from our drone testbed, corresponding to the data presented in Figs. \ref{D20}, \ref{D21}, and \ref{D22}. The proposed algorithm, MAVShield, leads to only a slight increase of $2.94 \, \%$ in CPU utilization compared to unencrypted MAVLink. Recall that the MAVShield encryption algorithm incorporates three additional operations— word splitting, S-box substitution, and XOR operations— on top of two Speck 64/128 round functions. This combination results in a modestly higher percentage increase in CPU usage compared to Speck 128/128. In contrast, the other five encryption algorithms exhibit significant increases in CPU utilization, which shows the efficacy of MAVShield.
In terms of memory consumption, MAVShield results in a slight increase of only $0.024 \, \%$, which is much lower than that of all the other encryption algorithms.  In terms of battery power consumption, there is only  a $5.653 \, \%$ increase in usage under MAVShield relative to unencrypted MAVLink, which indicates a high saving in battery power. This increase is notably low compared to that under the other encryption algorithms. In summary, our proposed cipher, MAVShield, stands out as the top-performing algorithm and a highly effective approach to encryption.

\begin{table*}[ht]
\caption{The figure shows the experimental results from our drone testbed. \iffalse{\bf (DONE: This table spills into the margins. Make it fit within the page width.)}\fi}
\label{table:III}
\resizebox{\textwidth}{!}{
\begin{tabular}{|l|l|l|l|l|l|l|}
\hline
\rowcolor[HTML]{D9E1F2} 
\multicolumn{1}{|c|}{\cellcolor[HTML]{D9E1F2}{\color[HTML]{330001} \textbf{Algorithm}}} & \multicolumn{1}{c|}{\cellcolor[HTML]{D9E1F2}\textbf{\begin{tabular}[c]{@{}c@{}}Battery Power\\ Consumption (mW)\end{tabular}}} & \multicolumn{1}{c|}{\cellcolor[HTML]{D9E1F2}\textbf{\begin{tabular}[c]{@{}c@{}}Memory \\ Available (B)\end{tabular}}} & \multicolumn{1}{c|}{\cellcolor[HTML]{D9E1F2}\textbf{\begin{tabular}[c]{@{}c@{}}CPU \\ Usage (\%)\end{tabular}}} & \multicolumn{1}{c|}{\cellcolor[HTML]{D9E1F2}{\color[HTML]{330001} \textbf{\begin{tabular}[c]{@{}c@{}}Percentage Increase in \\ Battery Power Consumption\end{tabular}}}} & \textbf{\begin{tabular}[c]{@{}l@{}}Percentage Increase in \\ Memory Consumption\end{tabular}} & \multicolumn{1}{c|}{\cellcolor[HTML]{D9E1F2}\textbf{\begin{tabular}[c]{@{}c@{}}Percentage Increase in \\ CPU Usage\end{tabular}}} \\ \hline
Unencrypted                                                                             & 13.620                                                                                                                          & 514819                                                                                                                & 33.70                                                                                                           &  -                                                                                                                                                                        &           -                                                                                    &   -                                                                                                                                \\ \hline
\cellcolor[HTML]{FFFFFF}{\color[HTML]{330001} MAVShield}                                & 14.390                                                                                                                          & 514695                                                                                                                & 34.69                                                                                                           & \cellcolor[HTML]{FFFFFF}{\color[HTML]{330001} 05.653}                                                                                                               & 0.024                                                                                   & 2.937                                                                                                                        \\ \hline
Speck 128/128                                                                           & 23.821                                                                                                                         & 513806                                                                                                                & 33.79                                                                                                           & 74.897                                                                                                                                                              & 0.196                                                                                   &  0.267                                                                                                                       \\ \hline
Speck 128/192                                                                           & 21.337                                                                                                                         & 513808                                                                                                                & 37.52                                                                                                           & 56.659                                                                                                                                                              & 0.196                                                                                   & 11.335                                                                                                                       \\ \hline
Speck 128/256                                                                           & 22.747                                                                                                                         & 513806                                                                                                                & 34.94                                                                                                           & 67.012                                                                                                                                                              & 0.196                                                                                   & 3.679                                                                                                                       \\ \hline
AES 128/128                                                                             & 20.386                                                                                                                         & 508888                                                                                                                & 35.70                                                                                                            & 49.676                                                                                                                                                              & 1.152                                                                                   & 5.934                                                                                                                       \\ \hline
ChaCha20                                                                                & 21.428                                                                                                                         & 513808                                                                                                                & 36.12                                                                                                           & 57.327                                                                                                                                                              & 0.196                                                                                   & 7.181                                                                                                                       \\ \hline
Rabbit                                                                                  & 23.573                                                                                                                         & 513800                                                                                                                & 34.80                                                                                                            & 73.076                                                                                                                                                               & 0.197                                                                                   & 3.264                                                                                                                       \\ \hline
\end{tabular}
}
\end{table*}

\section{Conclusions And Future Work}\label{Section12_con}
We integrated various existing encryption algorithms, viz., AES-CTR, ChaCha20, Speck-CTR, and Rabbit, into MAVLink. We proposed a novel cipher, MAVShield, designed to safeguard MAVLink-based communications. Also, we performed a security analysis of MAVShield, which includes a study of 24 distinct attacks on the proposed cipher using the NIST and Diehard test suites. Our analysis demonstrates the robust resistance of MAVShield
to differential cryptanalysis. Also, we thoroughly evaluated the performance of all five algorithms, viz., AES-CTR, ChaCha20, Speck-CTR, Rabbit, and MAVShield, and compared it with that of the standard unencrypted MAVLink protocol in terms of various metrics such as memory usage, battery power consumption, and CPU load, using a real drone testbed. Our performance evaluation demonstrates that MAVShield outperforms all the other encryption algorithms, and hence is well-suited for enhancing the communication link security of MAVLink, while achieving high performance.  In summary, MAVShield is a secure and efficient solution for protecting MAVLink-based communications in real-world deployments.
 
A direction for future research is to explore the algebraic cryptanalysis of MAVShield. Another open problem is to design a key-exchange protocol for UAV-GCS communications. 

\bibliographystyle{IEEEtran}
\bibliography{references}

\begin{thebibliography}{10}
\providecommand{\url}[1]{#1}
\csname url@samestyle\endcsname
\providecommand{\newblock}{\relax}
\providecommand{\bibinfo}[2]{#2}
\providecommand{\BIBentrySTDinterwordspacing}{\spaceskip=0pt\relax}
\providecommand{\BIBentryALTinterwordstretchfactor}{4}
\providecommand{\BIBentryALTinterwordspacing}{\spaceskip=\fontdimen2\font plus
\BIBentryALTinterwordstretchfactor\fontdimen3\font minus \fontdimen4\font\relax}
\providecommand{\BIBforeignlanguage}[2]{{%
\expandafter\ifx\csname l@#1\endcsname\relax
\typeout{** WARNING: IEEEtran.bst: No hyphenation pattern has been}%
\typeout{** loaded for the language `#1'. Using the pattern for}%
\typeout{** the default language instead.}%
\else
\language=\csname l@#1\endcsname
\fi
#2}}
\providecommand{\BIBdecl}{\relax}
\BIBdecl

\bibitem{adheebaT}
A.~Thahsin, A.~Ananthapadmanabhan, S.~Pathak, A.~Maity, and G.~S. Kasbekar, ``{Enhancing MAVLink Security: Implementation And Performance Evaluation Of Encryption On A Drone Testbed},'' in \emph{ICOIN 2025, The 39th International Conference on Information Networking}, 2025.

\bibitem{chao2010autopilots}
H.~Chao, Y.~Cao, and Y.~Chen, ``{Autopilots For Small Unmanned Aerial Vehicles: A Survey},'' \emph{International Journal of Control, Automation and Systems}, vol.~8, pp. 36--44, 2010.

\bibitem{pizzolante2023improving}
R.~Pizzolante, A.~Castiglione, F.~Palmieri, A.~Passaro, R.~Zaccagnino, and S.~La~Vecchia, ``{Improving Drone Security In Smart Cities Via Lightweight Cryptography},'' in \emph{International Conference on Computational Science and Its Applications}.\hskip 1em plus 0.5em minus 0.4em\relax Springer, 2023, pp. 99--115.

\bibitem{koubaa2019micro}
A.~Koub{\^a}a, A.~Allouch, M.~Alajlan, Y.~Javed, A.~Belghith, and M.~Khalgui, ``{Micro Air Vehicle Link (MAVLink) In A Nutshell: A Survey},'' \emph{IEEE Access}, vol.~7, pp. 87\,658--87\,680, 2019.

\bibitem{mavlinkio}
``{MAVLink Developer Guide}" {[ONLINE]},'' \url{https://mavlink.io/en/}, Dec. 2024.

\bibitem{khan2022secure}
N.~A. Khan, N.~Jhanjhi, S.~N. Brohi, A.~A. Almazroi, and A.~A. Almazroi, ``{A Secure Communication Protocol For Unmanned Aerial Vehicles},'' \emph{Cmc-computers Materials \& Continua}, vol.~70, no.~1, pp. 601--618, 2022.

\bibitem{nvdwebsite}
``{National Vulnerability Database- CVE-2020-10282 Detail}" {[ONLINE]},'' \url{https://nvd.nist.gov/vuln/detail/CVE-2020-10283}, Dec. 2024.

\bibitem{msgsign}
``{Message Signing (Authentication): Signature}" {[ONLINE]},'' \url{https://mavlink.io/en/guide/message_signing.html"}, Dec. 2024.

\bibitem{Xu}
H.~Xu, H.~Zhang, J.~Sun, W.~Xu, W.~Wang, H.~Li, and J.~Zhang, ``{Experimental Analysis of MAVLink Protocol Vulnerability on UAVs Security Experiment Platform},'' in \emph{2021 3rd International Conference on Industrial Artificial Intelligence (IAI)}, 2021, pp. 1--6.

\bibitem{hamza2024mavlink}
M.~A. Hamza, M.~Mohsin, M.~Khalil, and S.~M. K.~A. Kazmi, ``{MAVLink Protocol: A Survey Of Security Threats and Countermeasures},'' in \emph{2024 4th International Conference on Digital Futures and Transformative Technologies (ICoDT2)}.\hskip 1em plus 0.5em minus 0.4em\relax IEEE, 2024, pp. 1--8.

\bibitem{yustiarini2022comparative}
B.~Y. Yustiarini, F.~Dewanta, and H.~H. Nuha, ``{A Comparative Method For Securing Internet Of Things (IoT) Devices: {AES vs Simon-Speck} Encryptions},'' in \emph{2022 1st International Conference on Information System \& Information Technology (ICISIT)}.\hskip 1em plus 0.5em minus 0.4em\relax IEEE, 2022, pp. 392--396.

\bibitem{sabuwala2024approach}
N.~A. Sabuwala and R.~D. Daruwala, ``{An Approach To Enhance The Security Of Unmanned Aerial Vehicles (UAVs)},'' \emph{The Journal of Supercomputing}, vol.~80, no.~7, pp. 9609--9639, 2024.

\bibitem{beaulieu2017notes}
R.~Beaulieu, D.~Shors, J.~Smith, S.~Treatman-Clark, B.~Weeks, and L.~Wingers, ``{Notes On The Design And Analysis Of Simon And Speck},'' \emph{Cryptology ePrint Archive}, 2017.

\bibitem{boesgaard2008rabbit}
M.~Boesgaard, M.~Vesterager, and E.~Zenner, ``{The Rabbit Stream Cipher},'' \emph{New Stream Cipher Designs: The eSTREAM Finalists}, pp. 69--83, 2008.

\bibitem{knuthwebsite}
``{Q Ground Control (QGC) User Guide}" {[ONLINE]},'' \url{https://docs.qgroundcontrol.com/master/en/qgc-user-guide/index.html}, 2024.

\bibitem{ardu}
``{Autopilot}" {[ONLINE]},'' \url{"https://ardupilot.org/dev/docs/building-setup-linux.html"}, Dec. 2024.

\bibitem{cube}
\BIBentryALTinterwordspacing
{Cube Orange Plus Flight Controller}. [Online]. Available: \url{https://docs.px4.io/main/en/flight_controller/cubepilot_cube_orangeplus.html}
\BIBentrySTDinterwordspacing

\bibitem{RC}
\BIBentryALTinterwordspacing
{Boxer Radio Controller (M2)}. [Online]. Available: \url{https://www.radiomasterrc.com/products/boxer-radio-controller-m2}
\BIBentrySTDinterwordspacing

\bibitem{rukhin2001statistical}
A.~Rukhin, J.~Soto, J.~Nechvatal, M.~Smid, E.~Barker, S.~Leigh, M.~Levenson, M.~Vangel, D.~Banks, A.~Heckert \emph{et~al.}, \emph{{A Statistical Test Suite For Random and Pseudorandom Number Generators For Cryptographic Applications}}.\hskip 1em plus 0.5em minus 0.4em\relax US Department of Commerce, 2001, vol.~22.

\bibitem{bogos2022remark}
C.-E. Bogos, R.~Mocanu, and E.~Simion, ``{A Remark On NIST SP 800-22 Serial Test},'' \emph{Cryptology ePrint Archive}, 2022.

\bibitem{biham1991differential}
E.~Biham and A.~Shamir, ``{Differential Cryptanalysis of DES-Like Cryptosystems},'' \emph{Journal of CRYPTOLOGY}, vol.~4, pp. 3--72, 1991.

\bibitem{allouch2019mavsec}
A.~Allouch, O.~Cheikhrouhou, A.~Koub{\^a}a, M.~Khalgui, and T.~Abbes, ``{MAVSec: Securing The MAVLink Protocol For Ardupilot/PX4 Unmanned Aerial Systems},'' in \emph{2019 15th International Wireless Communications \& Mobile Computing Conference (IWCMC)}.\hskip 1em plus 0.5em minus 0.4em\relax IEEE, 2019, pp. 621--628.

\bibitem{kassim2022dmav}
G.~E. Kassim and S.~H. Hashem, ``{DMAV: Enhanced MAVLink Protocol Using Dynamic DNA Coding for Unmanned Aerial Vehicles},'' \emph{International Journal of Online \& Biomedical Engineering}, vol.~18, no.~11, 2022.

\bibitem{sabuwala2022securing}
N.~Sabuwala and R.~D. Daruwala, ``{Securing Unmanned Aerial Vehicles By Encrypting MAVLink Protocol},'' in \emph{2022 IEEE Bombay Section Signature Conference (IBSSC)}.\hskip 1em plus 0.5em minus 0.4em\relax IEEE, 2022, pp. 1--6.

\bibitem{bernstein2008chacha}
D.~J. Bernstein \emph{et~al.}, ``{ChaCha, A Variant Of Salsa20},'' in \emph{Workshop record of SASC}, vol.~8, no.~1.\hskip 1em plus 0.5em minus 0.4em\relax Citeseer, 2008, pp. 3--5.

\bibitem{manesh2019cyber}
M.~R. Manesh and N.~Kaabouch, ``{Cyber-Attacks On Unmanned Aerial System Networks: Detection, Countermeasure, And Future Research Directions},'' \emph{Computers \& Security}, vol.~85, pp. 386--401, 2019.

\bibitem{fanfakh2024parallel}
A.~Fanfakh, N.~Abduljalil, and A.~K.~M. Al-Qurabat, ``{Parallel Multi-Core Implementation Of The Optimized Speck Cipher},'' \emph{International Journal of Safety \& Security Engineering}, vol.~14, no.~3, 2024.

\bibitem{kurose2005computer}
J.~F. Kurose and K.~Ross, \emph{Computer Networking: A Top-Down Approach, 8/E}.\hskip 1em plus 0.5em minus 0.4em\relax Pearson Education India, 2021.

\bibitem{perlman2016network}
C.~Kaufman, R.~Perlman, M.~Speciner, and R.~Perlner, \emph{{Network Security: Private Communication In A Public World, 3/E}}.\hskip 1em plus 0.5em minus 0.4em\relax Pearson Education, Hoboken, NJ, 2022.

\bibitem{kwon2018empirical}
Y.-M. Kwon, J.~Yu, B.-M. Cho, Y.~Eun, and K.-J. Park, ``{Empirical Analysis Of {MAVLink} Protocol Vulnerability For Attacking Unmanned Aerial Vehicles},'' \emph{IEEE Access}, vol.~6, pp. 43\,203--43\,212, 2018.

\bibitem{beaulieu2015simon}
R.~Beaulieu, D.~Shors, J.~Smith, S.~Clark, B.~Weeks, and L.~Wingers, ``{The Simon and Speck Lightweight Block Ciphers},'' in \emph{Proceedings of the 52nd annual design automation conference}, 2015, pp. 1--6.

\bibitem{sawant2019implementation}
A.~G. Sawant, S.~Kamthe, Y.~Shaha, B.~Morajkar, and A.~Sakpal, ``{Implementation Of Simon \& Speck Algorithm},'' \emph{Journal of emerging technologies and innovative research}, vol.~6, no.~1, pp. 292--296, 2019.

\bibitem{boesgaard2003rabbit}
M.~Boesgaard, M.~Vesterager, T.~Pedersen, J.~Christiansen, and O.~Scavenius, ``{Rabbit: A New High-Performance Stream Cipher}.''\hskip 1em plus 0.5em minus 0.4em\relax Springer, 2003, pp. 307--329.

\bibitem{sanders2017practical}
C.~Sanders, \emph{{Practical Packet Analysis: Using Wireshark To Solve Real-World Network Problems}}.\hskip 1em plus 0.5em minus 0.4em\relax No Starch Press, 2017.

\bibitem{albrecht2009algebraic}
M.~Albrecht and C.~Cid, ``{Algebraic Techniques In Differential Cryptanalysis},'' in \emph{International Workshop on Fast Software Encryption}.\hskip 1em plus 0.5em minus 0.4em\relax Springer, 2009, pp. 193--208.

\bibitem{pervushin2021quantum}
B.~E. Pervushin, M.~Fadeev, A.~Zinovev, R.~K. Goncharov, A.~A. Santev, A.~E. Ivanova, and E.~O. Samsonov, ``{Quantum Random Number Generator Using Vacuum Fluctuations},'' \emph{Nanosystems: physics, chemistry, mathematics}, vol.~12, no.~2, pp. 156--160, 2021.

\bibitem{freqtest}
``{Frequency Test}" {[ONLINE]},'' \url{https://www.itl.nist.gov/div898/software/dataplot/refman1/auxillar/freqtest.htm}, Jan. 2025.

\bibitem{li2020parallel}
K.~Li, J.~Zhang, P.~Li, A.~Wang, and Y.~Wang, ``{Parallel Implementation Of The Non-Overlapping Template Matching Test Using CUDA},'' \emph{China Communications}, vol.~17, no.~8, pp. 234--241, 2020.

\bibitem{diehard}
T.~N. Van, ``{Diehard Statistical Tests Results on GINARs RNG},'' \url{https://github.com/GINARTeam/Diehard-statistical-test/blob/master/Diehard_Report.pdf}, 2019.

\bibitem{telemetry}
\BIBentryALTinterwordspacing
{SiK Telemetry Radio V3}. [Online]. Available: \url{https://holybro.com/collections/telemetry-radios}
\BIBentrySTDinterwordspacing

\bibitem{GPS}
\BIBentryALTinterwordspacing
{HEX Here3+ CAN GNSS GPS Module with iStand}. [Online]. Available: \url{https://robu.in/product/hex-here3-can-gnss-gps-module-with-istand/}
\BIBentrySTDinterwordspacing

\bibitem{LiPo}
\BIBentryALTinterwordspacing
{10000mAH 6S 25C 22.2V LiPo Battery}. [Online]. Available: \url{https://robokits.co.in/batteries-chargers/drone-batteries}
\BIBentrySTDinterwordspacing

\bibitem{motor}
\BIBentryALTinterwordspacing
{MN4006 Antigravity Type 4-6S UAV Motor}. [Online]. Available: \url{https://store.tmotor.com/categorys/multi-rotor-drone-motor}
\BIBentrySTDinterwordspacing

\end{thebibliography}

\end{document}